\documentclass[12pt]{article}
\usepackage{amsfonts}
\usepackage{bm}
\usepackage{amsmath}
\usepackage{epsfig}

\usepackage{amssymb,lscape}
\usepackage[matrix,arrow,curve]{xy}

\newcommand{\bmat}{\left(\begin{array}}
\newcommand{\emat}{\end{array}\right)}

\def\gtrsim{\mathrel{\raise.3ex\hbox{$>$\kern-.75em\lower1ex\hbox{$\sim$}}}}

\def\ti{\tilde}
\def\g{\gamma}

\def\-{\hphantom{-}}

\def\s2{\frac{1}{\sqrt2}}

\def\T{{\rm T}}
\def\Z{{\mathbb Z}}

\def\mg{m_{3/2}}
\def\mg2{m^2_{3/2}}

\def\Dsl{\,\raise.15ex\hbox{/}\mkern-13.5mu D} 
\def\rep#1{\mbox{{\bf #1}}}

\def\aneq{\not=}

\def\beq{\begin{equation}}
\def\eeq{\end{equation}}
\def\beqa{\begin{eqnarray}}
\def\eeqa{\end{eqnarray}}
\newcommand{\bbR}{\mathbb{R}}

\DeclareMathOperator{\SL}{\mathit{SL}}

\DeclareMathOperator{\E7}{\mathit{E}_{7}}

\newcommand{\Tsub}{O(6,6) \times \SL(2,\bbR)}

\newcommand{\nn}{\nonumber}

\begin{document}

\pagestyle{plain}

\makeatletter
\@addtoreset{equation}{section}
\makeatother
\renewcommand{\theequation}{\thesection.\arabic{equation}}
\pagestyle{empty} \rightline{IPhT-T11/014}
\begin{center}
\LARGE{On Type IIB moduli stabilization and \\${\cal N}=4, 8$
supergravities\\[10mm]}
\large{ Gerardo Aldazabal${}^{a,b}$, Diego Marqu\'es$^c$,  Carmen
N\'u\~nez$^d$ and \\Jos\'e  A.
Rosabal${}^a$
 \\[6mm]}
\small{
${}^a${\em Centro At\'omico Bariloche, ${}^b$Instituto Balseiro
(CNEA-UNC) and CONICET} \\[-0.3em]
8400 S.C. de Bariloche, Argentina.\\
${}^c$ {\em Institut de Physique Th\'eorique,
CEA/ Saclay} \\
91191 Gif-sur-Yvette Cedex, France  \\
${}^d${\small\it  Instituto de Astronom\'ia y F\'isica del Espacio
(CONICET-UBA) and\\
Departamento de F\' isica, FCEN, Universidad de Buenos Aires}\\
C.C. 67 - Suc. 28, 1428 Buenos Aires, Argentina\\[1cm]}
\small{\bf Abstract} \\[0.5cm]\end{center}
We analyze $D=4$  compactifications of Type IIB theory with generic,
geometric and non-geometric, dual fluxes turned on. In particular,
we study  ${\cal N}=1$ toroidal orbifold compactifications that
admit an embedding of the untwisted sector into gauged ${\cal N}=4,
\ 8$ supergravities. Truncations, spontaneous
breaking of supersymmetry and the inclusion of sources
 are discussed. The algebraic identities satisfied by the
supergravity gaugings are used to implement the full set of
consistency constraints on the background fluxes. This allows to
perform a generic study of ${\cal N}=1$ vacua and identify large
regions of the parameter space that do not admit complete
moduli stabilization.
Illustrative examples of AdS and Minkowski vacua are presented.

{\small}

\newpage
\setcounter{page}{1}
\pagestyle{plain}
\renewcommand{\thefootnote}{\arabic{footnote}}
\setcounter{footnote}{0}

\tableofcontents

\section{Introduction}
\label{seci}

Compactifications of superstring theory to four dimensions must be addressed
if a link between string theory and the physics of the
observable world is to be established.
From a phenomenological approach, the compactification process must fulfill some
very basic requirements. For instance, part of the
original supercharges must be projected out in order to obtain a
$D=4$
phenomenologically acceptable ${\cal N}=1$ supersymmetric  (or ${\cal N}=0$)
theory and a chiral fermionic spectrum must be allowed. These
requirements
are linked to the structure of the internal manifold and to the presence
of sources
like D-branes or orientifold planes.  Clearly, other kinds of compactifications
with more supersymmetry charges are also worth being studied  in order
to explore other
aspects of the theory like, for example, the AdS/CFT correspondence.

In a generic compactification, field strength background fluxes for
the different form fields can be turned on. An appealing feature of
these so called  flux compactifications \cite{Lectures, samt}  is
that scalar fields, which would be moduli in the absence of fluxes,
acquire a potential \cite{gvw} and thus, vacuum degeneracy can be
lifted. Unfortunately, a string theory formulation of
compactifications with fluxes is not yet available and we must
tackle them by using the low energy effective supergravity theory.
More precisely, superstring flux compactifications would be
described by gauged supergravities \cite{deWit:2002vt}-\cite{Weidner:2006rp}, namely, deformations of ordinary
abelian supergravity theories where the deformation parameters
(gaugings) correspond to the quantized fluxes
\cite{Villadoro:2005cu}-\cite{Dall'Agata:2009gv}. Consistency constraints
on gaugings are needed for the supergravity theory to be well
defined \cite{samt}.

The connection between a string compactification and the gauged
supergravity effective theory is not fully evident and calls for
some interpretation. In fact, a first sight inspection seems to
indicate  that there are many more gaugings available than
background fluxes.  For instance, an orientifold compactification
with fluxes  of the Type IIB supergravity action (the low energy
effective theory of  Type IIB superstrings), leads to a  $D=4$,
${\cal N}=4$ supergravity theory  involving only a small subset of
all possible gaugings (identified with Type IIB 3-form fluxes).
An explanation of this missmatch relies in the fact  that,  starting with the
effective
$D=10$ theory, one is missing some key stringy ingredients like, for
instance, T-duality invariance. Actually,  if different dualities
(like T-duality,  IIB
S-duality, M-theory or heterotic/Type I S-duality),
expected from the underlying string theory,
 are enforced
in the low energy four dimensional action,  new
``dual fluxes'' must be invoked, and a one to one correspondence
between gaugings and fluxes can be established \cite{stw,acfi}.
 A way to accomplish
this correspondence is to compare the (orbifold projected) gauged
supergravity algebra with the duality completion of the algebra
satisfied by  compactified $D = 10$ vector bosons (see
\cite{acr,Dall'Agata:2009gv}). Interestingly enough, the Jacobi
identities (JI) of the resulting algebra encode the constraints that
dual fluxes  must obey, including and generalizing the already known
Bianchi identities and tadpole cancelation equations. In this
interpretation, the four dimensional supergravity theory integrates
information about the stringy aspects of the starting configuration
and it is not just the reduction of a ten dimensional effective
supergravity action.

It must be recalled, however,  that even if this correspondence is
established, the  interpretation in terms of $D = 10$ fields is
subtle. In particular, non geometric fluxes must be incorporated,
meaning that globally  matching solutions
requires,
besides diffeomorphisms and gauge invariance, T or S-duality (or
generically U-duality) transformations as part of the transition
functions.
The concept
of Generalized Geometry has been proposed as an appealing framework
to describe flux compactifications \cite{GMPT}.

Here we study ${\cal N}=1$ supergravity models resulting
from orbifold compactification with fluxes. Our starting point is
the  framework of \cite{acr} where Type IIB orientifold
compactifications plus a $\mathbb{Z}_2\times\mathbb{Z}_2$ orbifold
projection ensuring tori factorization are considered. The goal of
the paper is twofold. On the one hand we analyze the conditions
under which the untwisted sectors of such string theories correspond
to truncations of gauged supergravities. These consistency
conditions are then used to implement the full set of constraints on
fluxes. Some of these constraints
have not received much attention previously.
We finally solve all the consistency
equations for fluxes and analyze their impact in
the analysis of moduli fixing.

In the absence of
sources, the untwisted sector of such compactification
is just the orbifold projection of an underlying
${\cal N}=4$  gauged supergravity theory. This sector was first found in
\cite{acfi}, were it was shown that $2^7$ dual fluxes are needed to
ensure duality invariance. In \cite{acr}, a subset of $2^6$ fluxes
was shown to describe globally non-geometric compactifications of
F-theory (which admit a geometric local description in terms of $D =
10$ supergravity) and a consistent procedure for incorporating
F-theory seven branes into the gauge algebra  was discussed.
Moreover, the algebra of gauge generators and JI  leading to
constraints on fluxes  were derived, and shown to coincide with that
of
${\cal N} = 4$ gauged supergravity, when there are no sources.

In Section \ref{sec0}, we start introducing some essential
features of Type IIB/O3 orientifolds,
 their moduli space and effective action in the presence of
the complete set of fluxes allowed by the orbifold.

We begin Section \ref{sec1} with a review of basic notions on gauged ${\cal N} =
4$ supergravity, relevant for our discussion. We then analyze the ``shift
matrices'' and
the  scalar potential of gauged ${\cal N}=4$ supergravity. We
discuss the projection to ${\cal N}=1$ and a scheme for spontaneous
supersymmetry breaking solutions. We also consider the possibility
of embedding the model into an underlying ${\cal N}=8$ gauged
supergravity theory, and derive new constraints on dual fluxes. The analysis is
completed
 with a discussion about the incorporation of
sources. The full analysis allows us to understand the embedding of the string
effective action in gauged supergravity and the link between fields, parameters,
constraints, etc.

Section \ref{sec4} is devoted to a generic analysis of moduli fixing
in Type IIB orbifolds (possibly including
 three and seven branes). The complete set of physical and algebraic
constraints on
moduli and fluxes is
 reviewed and an exclusion principle allowing to identify  models that
do not admit full moduli stabilization is formulated. We use this
result to determine large regions of the parameter  space that are
excluded, and focus on models with potentiality to fully stabilize
all moduli. Illustrative examples of SUSY vacua are presented.

Finally, Section \ref{conc} presents some conclusions and an
outlook. We collect in the Appendices some useful equations and
 tables containing fluxes and their transformation
properties.

\section{Type IIB/O3 ${\cal N}=1$ compactifications with generic
fluxes} \label{sec0}
We  consider Type IIB/O3 orientifold compactifications on
$T^6/[\Omega_P (-1)^{F_L} \sigma\times \Gamma]$, where $\Omega_P$ is
the worldsheet parity operator, $(-1)^{F_L}$ is the space-time
fermionic number for left-movers and $\sigma$ is an involution
operator acting on internal coordinates as $\sigma(x^i) = - x^i$.
$\Gamma$ is an orbifold projection to be specified below. We
introduce a basis of 3-forms for this space
\beqa && \alpha_0 = dx^1 \wedge dx^2 \wedge dx^3 \ , \ \ \ \ \ \
\alpha_i^{\ j+3} = \frac{1}{2} \epsilon_{ilm} dx^l\wedge dx^m \wedge
dx^{j+3}\nn \, ,
\\
&&\beta^0 = dx^4 \wedge dx^5 \wedge dx^6 \ , \ \ \ \ \ \ \beta^i _{\
j+3} = - \frac{1}{2} \epsilon_{jlm} dx^{l+3}\wedge dx^{m+3} \wedge
dx^{i} \, ,\eeqa
with elements normalized according to $\int dx^1\wedge \dots \wedge
dx^6 = 1$.  In addition, we take the basis of closed 2-forms and
their $\ast_6$-dual 4-forms as
\beq \omega_{i j} = - dx^i \wedge dx^{j+3} \ , \ \ \ \ \ \ \tilde
\omega_{i j} = \frac{1}{4}\ \epsilon_{ilm}\epsilon_{jpq}\ dx^{l}
\wedge dx^{p + 3} \wedge dx^{m} \wedge dx^{q + 3} \, .\eeq
We define a complex structure whose deformations are controlled by
complex structure  moduli $U^m_{\ \ n}$
\beq \Omega = dz^1 \wedge dz^2 \wedge dz^3 \ , \ \ \ \ dz^m = dx^m +
i U^m_{\ \ p} dx^{p+3} \, ,\eeq
and a complexified K\"ahler 4-form with K\"ahler  moduli $T^{ij}$
\beq {\cal J}_c \equiv C_4 + \frac{i}{2} e^{-\phi} J \wedge J = i
T^{ij} \tilde \omega_{i j} \ . \eeq
Finally, the RR axion and the string coupling combine into the
axion-dilaton modulus as \beq S = e^{-\phi} + i C_0 \, .\eeq
The $1+9+9$ complex moduli span a K\"ahler manifold with K\"ahler
potential \cite{gl}
\beqa K &=& - \log(S+\bar S) - \log \left[-i \int \Omega \wedge \bar
\Omega\right] - 2 \log\left[\frac{1}{6}\int J\wedge J \wedge
J\right] \nn\\ && \nn\\&=& -\log[(S + \bar S) \det(U+\bar U) \det
(T+\bar T)] \, .\label{PotKahler}\eeqa

For trivial $\Gamma$, this orientifold leads to ${\cal N} = 4$
effective theories in $D=4$ (see \cite{aacg}). The large amount of
SUSY can be broken in particular to ${\cal N} = 1$ by orbifolding
the theory with a discrete symmetry group $\Gamma \subset SU(3)$. In
this case the structure group of the tangent bundle is still the
trivial one, but it is enhanced to $SU(3)$ at the orbifold fixed
points. Then, two different sectors of states can be distinguished.
The untwisted sector is given by direct truncation of the parent
${\cal N}=4$ theory and its $SL(2,\mathbb{Z}) \times SO(6,6;
\mathbb{Z})$ gauge group, while the twisted states localized at the
orbifold singularities transform under a larger symmetry group. In
this work we restrict to the untwisted sector. We consider $\Gamma =
\mathbb{Z}_2 \times \mathbb{Z}_2$, which orbifolds the internal
space as
\begin{center}
\begin{tabular}{c|c|c|c|c|c|c}
$ $ & $x^1$ & $x^2$ & $x^3$ & $x^4$ & $x^5$ & $x^6$ \\
\hline
$\mathbb{Z}_ 2$ & $-$ & $-$ & $+$ & $-$ & $-$ & $+$ \\
$\mathbb{Z}_2$ & $+$ & $-$ & $-$ & $+$ & $-$ & $-$
\label{z2z2}
\end{tabular}
\end{center}
leaving only the diagonal K\"ahler and complex moduli $T_i = T^{ii}$
and $U_i = U^i_{\ i}$, which leads to
\beq K = - \log (S + \bar S ) - \sum_{i = 1}^3 \log (U_i + \bar U_i)
- \sum_{i = 1}^3 \log (T_i + \bar T_i)\, .\eeq
After orbifoldization, the global symmetry group associated to the
moduli space of the internal manifold is $SL(2,\mathbb{Z})^7$, where
each $SL(2,\mathbb{Z})$ generates modular transformations for each
of the seven moduli $S$, $U_i$ and $T_i$. These fields can be
grouped in the vector representation $ \mathbb{T} \equiv \mathbf{7}
= (S; T_1, T_2, T_3; U_1, U_2, U_3)$ and the flux parameters
transform in the spinorial representation $\mathbb{G} \equiv
\mathbf{128}$ \cite{acfi}, each flux being characterized by a given
set of  Weyl spinors $(\pm,\pm,\pm,\pm,\pm,\pm,\pm)$ (see \cite{acr}
for details on the spinor formalism).

  In the
absence of fluxes, any  moduli configuration  corresponds to a
possible vacuum in which all fields are massless. The orientifold
involution allows to turn on RR $F_3$ and NSNS $H_3$ background
fluxes, which lift the moduli space generating a superpotential
\cite{gvw} \beq W = \int (F_3 - i S H_3) \wedge \Omega \, .\eeq Clearly,
this superpotential cannot fully stabilize all moduli (in
particular, it is independent of $T_i$). However, string dualities
can be invoked to construct an effective four dimensional theory
in which $W$ depends on all the moduli,
being the low energy limit of a ten dimensional string set up,
rather than ten dimensional supergravity. To achieve this
commitment, one can promote T-duality and S-duality
 to symmetries of the four dimensional theory. For
example, in order to restore mirror symmetry between IIA/IIB
orientifold compactifications, new set of fluxes were introduced in
\cite{stw}-\cite{acfi}. T-duality was used to connect the $H_3$
fluxes with non-geometric fluxes through the chain
\begin{equation}
H_{mnp}{\stackrel{\T_m} {\longleftrightarrow}}-\omega^m_{np}{\stackrel{\T_n}
{\longleftrightarrow}}-Q^{mn}_p{\stackrel{\T_p}
{\longleftrightarrow}}R^{mnp}\ . \label{Tdualitychain}
\end{equation}
Some of these new objects were given a clear interpretation: the
geometric $\omega$ fluxes correspond to structure constants of
compactifications on twisted tori \cite{Scherk:1979zr} and the
$Q$-fluxes correspond to locally geometric but globally
non-geometric constructions \cite{kst}. The $R$ fluxes instead,
correspond to highly non-geometric compactifications, not even
admitting a local description \cite{stw}. Only the $H$ and $Q$
fluxes survive the orientifold projection and many of their
components vanish after modding by the $\Gamma$ symmetry. Under
S-duality, the axion-dilaton transforms as \beq S \to \frac{k S - i
\ell}{imS + n} \ , \ \ \ \ kn - lm = 1 \, ,\eeq and
the 3-form fluxes rotate as doublets \beq \left(\begin{matrix}F_3 \\
H_3
\end{matrix}\right) \rightarrow \left(\begin{matrix}k & \ell \\ m& n
\end{matrix}\right)\left(\begin{matrix}F_3 \\ H_3
\end{matrix}\right) \, .\eeq In \cite{acfi} the $Q$-fluxes were included in the superpotential, and in order
to achieve S-invariance, a new set of fluxes $P^{ij}_k$ was considered.
They both conform an $SL(2,\mathbb{Z})$-doublet \beq
\left(\begin{matrix}Q \\ P \end{matrix}\right) \rightarrow
\left(\begin{matrix}k & \ell \\ m & n
\end{matrix}\right)\left(\begin{matrix}Q \\ P \end{matrix}\right)\, .
\eeq Since the $SL(2,\mathbb{Z})$ symmetry relating these fluxes is
a symmetry of the Type IIB supergravity equations of motion, the
$P$-flux parameters are expected to correspond to compactifications
admitting a ten dimensional local supergravity description. The full
superpotential
is \beq W =
\int [(F_3  - i S H_3) + (Q - iSP) {\cal J}_c] \wedge \Omega \ , \ \
\ \ \ \  (Q{\cal J}_c)_{p_1p_2p_3} = \frac{1}{2} Q^{ab}_{[p_1}
({\cal J}_c)_{p_2p_3] ab}\  . \label{tf}\eeq
Finally, by further demanding
the four dimensional
effective theory to be invariant under $SL(2,\mathbb{Z})^7$, new
``primed'' fluxes must be incorporated \cite{acfi}. Then, the full
invariant (modulo K\"ahler transformations) superpotential reads
\beq W = \int (f_+ -i S f_-)\cdot e^{{\cal J}_c} \wedge\Omega \,
,\label{superpotgeneral}
 \eeq
 where $f_\pm$ denote the formal sums
 \beq
f_+ = F_{abc} + Q^{ab}_c + Q'^a_{bc} + \tilde F'^{abc} \ , \ \ \ \ \
\ f_- = H_{abc} + P^{ab}_c + P'^a_{bc} + \tilde H'^{abc}\, .
 \eeq
Tilded and untilded 3-forms are related as $
\tilde{F}^{ijk}\equiv \frac{1}{3!}\epsilon^{ijkopq}F_{opq}$.

Besides (\ref{tf}), we have defined the 3-forms
\beqa
(Q' \cdot {\cal J}_c^2)_{p_1 p_2p_3} &=&\frac{1}{32}
Q'^{a}_{[p_1p_2}({\cal J}_c)_{p_3] a i_1i_2}
({\cal J}_c)_{i_3i_4p_5p_6} \epsilon^{i_1i_2i_3i_4i_5i_6} \, ,\\
(\tilde F' \cdot{\cal J}_c^3)_{p_1 p_2p_3} &=& \frac{5}{128} \tilde
F'^{abc}({\cal J}_c)_{[p_1p_2p_3a} ({\cal J}_c)_{bc]i_1i_2}({\cal
J}_c)_{i_3i_4i_5i_6} \epsilon^{i_1i_2i_3i_4i_5i_6}\, .\eeqa
An expanded version of the superpotential in terms of flux
parameters can be found in Appendix A (we do not include
contributions from twisted sectors, which are also expected).

Although the effective potential that we have just motivated looks
rather generic, its parameters are strongly constrained by
consistency requirements. From a higher dimensional point of view
these constraints originate, for instance, in Bianchi identities and
anomaly cancelation conditions. An efficient way of finding such
constraints was proposed in \cite{acr} by constructing a full
invariant algebra through systematic application of $SL(2,\Z)^7$
dualities on the well known algebra of gauge generators associated
to non-geometric Type IIB fluxes. We summarize the results of \cite{acr} in
Appendix A, and extend them by including also ``primed'' fluxes.

For particular sets of fluxes, the untwisted scenario that we have
just described corresponds to a truncation of gauged ${\cal N}=4$
supergravity \cite{acr}. In the following section we show that this
holds in general, and stress the connection between orbifold
compactifications of string theory and gauged supergravity
truncations.

\section{Gauged ${\cal N}=4 , \ 8$ supergravity embeddings}
\label{sec1}

The full $SL(2,\mathbb{Z})^7$ duality group of the untwisted sector
of the ${\cal N}=1 $ theory discussed in the previous section is
just the orbifold projection of the $SL(2,\mathbb{Z}) \times SO(6,6;
\mathbb{Z})$ global symmetry group of gauged ${\cal N}=4$ supergravity.
Therefore
we expect the untwisted sector of the $D=4$ effective action to
correspond to a truncation of a  gauged  ${\cal N} = 4$
supergravity,
and the full set of fluxes to correspond to all possible gaugings in
\cite{stw}. Actually, this connection was analyzed
for a subset of
gaugings
of Type IIB in \cite{acr}  and
of Type IIA in \cite{Dall'Agata:2009gv}.
 Here we further elaborate and generalize this
connection to the full set of gaugings allowed by the orbifold.

We begin by reviewing gauged ${\cal N} = 4$ supergravity \cite{sw},
focusing the attention on the scalar sector, consistency constraints
and supersymmetry variations of the fermions. We analyze orbifold
truncations of these theories, and their link to orbifold
compactifications in string theory (with and without sources).
Spontaneous supersymmetry breaking is also discussed.

\subsection{Gauged  ${\cal N}=4$ supergravity truncations}

Let us start by collecting some known facts about  ${\cal N}=4$
gauged supergravity that are needed for our discussion (see \cite{sw} for
details). These theories have $1+9+9+...$ complex
scalars $S\ + \ U^m{}_{\tilde q}\ + \ T^{l\tilde m}\ +\ ...$\footnote{This
notation highlights the connection with the string
 compactification described in the previous section and the dots are associated
to the presence of extra vector multiplets when $n\ne 0$.} that
span the coset
\begin{equation}\frac{SL(2)|_B}{U(1)} \times
\frac{SO(6,6+n)}{SO(6)\times SO(6+n)}\, .\end{equation}

 The first factor in the coset can be parameterized by the metric
$M^{\alpha \beta}= {\rm Re}[{\cal V}^{\alpha}{\cal V}^{\beta *}] $ with

\begin{equation} {\cal V}^{\alpha}=
\frac{1}{\sqrt{s_x}} \left(\begin{matrix}1 \\
-i{S}\end{matrix}\right) \ , \ \ \ \ \ M^{\alpha \beta} =
\frac{1}{s_x} \left(\begin{matrix} 1 & {s_y} \\{s_y} &
|S|^2\end{matrix}\right) \, ,\quad \quad \label{msu2} \end{equation}
where $S=s_x+is_y$.
The second factor is characterized by the vielbeins ${\cal V}^a$ and
${\cal V}^m$ with $m=1,\dots, 6$ and $a=7,\dots, 12+n$ and the coset
can be parameterized by the metric
\begin{equation}
  M_{M N}=({\cal V}{\cal
V}^T)_{MN}\, ,
\label{md6}
\end{equation}
where capital indices take values $M,N,...=1,\dots, 12+ n$ such that
the first six indices run over the rank of the $SO(6)$ factor whereas
the other values run over the rank of the $SO(6+n)$ factor (in this paper
we will mainly restrict  to $n=0$).

Supersymmetry and
anomaly cancelation require  a potential for the
scalars living in the vector multiplets, namely
\begin{multline}
V_{\mathcal{N}=4}= \frac{1}{16}\left[f_{\alpha MNP}f_{\beta
QRS}M^{\alpha\beta}\left(\frac13
M^{MQ}M^{NR}M^{PS}+\left(\frac23\eta^{MQ}-M^{MQ}\right)\eta^{NR}\eta^{PS}
\right)\right.\\
\left.-\frac49 f_{\alpha MNP}f_{\beta
QRS}\epsilon^{\alpha\beta}M^{MNPQRS}+3\xi_\alpha^M\xi_\beta^NM^{\alpha\beta}M_{
MN}\right]\, ,\label{scag}
\end{multline}
together with a topological term for the vector fields \cite{sw}.
Here $M^{MNPQRS}$ is a completely antisymmetric tensor which can be
expressed in terms of the vielbeins as
\begin{equation} M^{MNPQRS} = \epsilon^{mnpqrs} {\cal V}_m^{\ \ M}{\cal V}_n^{\  N}{\cal V}_p^{\  P}{\cal V}_q^{\  Q}{\cal V}_r^{\  R}{\cal V}_s^{\  S}\, ,
\end{equation}
and the indices are lowered and
raised with the
  off diagonal  $SO(6,6)$ metric\footnote{Since
our metric is rotated with respect to $\eta^{M N}=diag(1,\dots ,1,-1,\dots ,-1)$ used in \cite{sw}, we take  the rotated
inverse vielbien ${\cal V}_M^{\quad P}\rightarrow B_M^{\quad N}{\cal V}_N^{\quad P}$, where  $B$ is given by
$B_N^{\quad M}=  \frac{1}{\sqrt{2}} \left( \begin{array}{cc} I_{6\times 6} &
I_{6\times 6}\\
I_{6\times 6} & -I_{6\times 6} \end{array} \right).$}

\begin{equation}
 \eta^{M N}=\eta_{M N}=  \left( \begin{array}{cc}
 0 & I_{6\times 6} \\
 I_{6 \times 6} & 0  \end{array} \right) \ , \ \ \ \ \ \ \ \eta^{M N}={\cal V}_P{}^{M}\,\eta^{PQ}\,{\cal V}_Q{}^{
 N}\ .
\end{equation}

The embedding tensors
$f_{\alpha MNP}$ and $ \xi_{\alpha M}$
with $\alpha=\pm 1/2$ and $M=1\ldots 12+n$, encode all possible
gaugings of $\mathcal{N}=4$ supergravity, with $n$ extra vector
multiplets of $SL(2,\mathbb{Z})\times SO(6,6+n;\mathbb{Z})$. For
$n=0$ they live in  $(\mathbf{2},\mathbf{220})$ and
$(\mathbf{2},\mathbf{12})$ representations, respectively. In what
follows,  we will not include the $\xi_{\alpha M}$ fluxes,
which are projected out by the orbifold.

Consistency requires the
gaugings  to satisfy several constraints that can be obtained from
the algebra of gauge generators, namely  \cite{sw}
\begin{equation}
\left[Z_{\alpha A}, Z_{\beta B}\right] = f^P_{\alpha AB }\ Z_{\beta
P} = f^P_{\beta AB }\ Z_{\alpha P}\, ,
 \label{AlgebraSW}
\end{equation}
 where the last equality is enforced to ensure the antisymmetry
of the commutator. The quadratic constraints imposed for algebraic
consistency are
\begin{eqnarray}
 f_{\alpha [AB}^P f_{\beta CD] P} = 0\, , \qquad
 \epsilon^{\alpha
\beta} f_{\alpha AB}^P f_{\beta CD P}  = 0\, .
\label{ji}
\end{eqnarray}
We will refer to the last set of equations as ``antisymmetry conditions''. We will see later that their imposition plays a crucial role in moduli fixing, a fact that was also noted in \cite{Dibi}.

In order to study  truncations or  spontaneous supersymmetry
breakings, it appears enlightening to look at the quadratic
terms involving fermions
in  the supergravity Lagrangian and the fermionic transformations
under supersymmetry. With this in mind, it  proves useful to work
with the $SU(4)$ covering group rather than with  $SO(6)$. Thus, in
particular, an  $SO(6)$ vector basis  $e^m$, $m=1,\dots 6$, can be
expressed  in terms of a set of antisymmetric matrices $(\gamma^m)^{ij}$,
$i,j=1,\dots,4$ (details are given in  Appendix B).  Closely
following \cite{sw} we write
\begin{equation}
 {\cal V}^{-1}{\cal L}=\dots\frac{1}{3}A_1^{ij}\overline{\psi}_{\mu
i}\Gamma^{\mu \nu}\psi_{\nu
j}-\frac{1}{3}iA_2^{ij}\overline{\psi}_{\mu i}\Gamma^{\mu}\chi_
j+iA_{2ai}{}^{j}\overline{\psi}_{\mu
i}\Gamma^{\mu}\lambda_j^a+h.c\dots
\end{equation}
with ${\psi}_{\mu}^i$, $\chi_ j$ and $\lambda_j^a$  the gravitini,
dilatini and gaugini, respectively. The  corresponding shift matrices
$A_1$, $A_2$ and $A_{2a}$ are given in terms of the gaugings and the
vielbeins ${\cal V}_m^{\ \ M}\mapsto ({\cal V})^{M ij}={\cal V}_m^{\
\  M}(\gamma^m)^{ij}$ by
\begin{eqnarray}
 A_1^{ij}&=&\epsilon^{\alpha \beta}\bar {\cal V}_{\alpha}({\cal
V})^{M i
k}(\overline{\cal V})^{N  k l}({\cal V})^{P l j}f_{\beta MNP}\, ,\\
 A_2^{ij}&=&\epsilon^{\alpha \beta}{\cal V}_{\alpha}({\cal V})^{M i
k}(\overline{\cal V})^{N  k l}({\cal V})^{P l j}f_{\beta MNP}\, ,\\
\overline{A}_{2ai}{}^{j}&=&-\epsilon^{\alpha \beta}{\cal
V}_{\alpha}({\cal V})_{a}^{\  M}({\cal V})^{N i k}(\bar{\cal
V})^{Pkj}f_{\beta MNP}\, .  \label{shiftmat}
\end{eqnarray}
They verify (upon use of the consistency constraints)
\begin{equation}
 -\frac{1}{4}\delta_j^iV=\frac{1}{3}A_1^{ik}\bar
A_{1jk}-\frac{1}{9}A_2^{ik}\bar A_{2jk}-\frac{1}{2}A_{2
aj}{}^{k}\bar A_{2 ai}{}^{k}\, .\label{massmatrix}
 \end{equation}
The (order $g$) supersymmetry transformations of the fermions read
 \begin{eqnarray}
 \delta\psi_{\mu}^i & = &
2D_{\mu}\epsilon^i-\frac{2}{3}A_1^{ij}\Gamma_{\mu}\epsilon_j\dots\, ,\\
\delta\chi^i & = & -\frac{4}{3}iA_2^{ij}\epsilon_j\dots\, ,\\
 \delta\lambda_a^i & = & -2iA_{2aj}^{\ \ \  i}\epsilon^j\dots\,
.\label{fermionvar}
 \end{eqnarray}

 Now we can proceed to the orbifold projection on the scalar and fermionic
sectors. The full   $\xi_{\alpha M}$ is projected out, and  the only
surviving gaugings  are
\beq f_{\alpha ABC} \ , \ \ \ \ \ \ \ A\  {\rm mod}\ 3 \neq B\ {\rm
mod}\ 3 \neq C\ {\rm mod}\ 3 \, . \eeq Additionally, non-diagonal moduli
are projected out, and only $S$, $T_i$ and $U_i$ parameterize the
moduli space. The metric matrix organizes into blocks associated to
each factorized torus, and reads

\begin{equation}
 M_i= \frac{1}{t_{ix}u_{ix}}\left(
\begin{array}{llll}
|T_i|^2  & -|T_i|^2u_{iy} & t_{iy}u_{iy} & t_{iy} \\\\
-|T_i|^2u_{iy} & |T_i|^2 |U_i|^2 & -t_{iy} |U_i|^2 &  -t_{iy} u_{iy}\\\\
 t_{iy} u_{iy} & -t_{iy} |U_i|^2 &|U_i|^2 & u_{iy} \\\\
 t_{iy} &  -t_{iy} u_{iy} &  u_{iy} & 1
\end{array}
\right)
\label{Mi}
\end{equation}
where we have written the non vanishing entries in positions
$M_i^{i+3k,i+3m}$ with $k,m=0,1,2,3$. Also,  $T_i=t_{ix}+ i t_{iy}$ and
$U_i=u_{ix}+ i u_{iy}$.\footnote{It is easy to see
that, after a reorganization of indices, the metric can be expressed
as $ M^{M N}={\rm diag}( M_{T_1}\otimes M_{U_1},M_{T_2}\otimes
M_{U_2},M_{T_3}\otimes M_{U_3} )$, where
\begin{eqnarray}
M_{U_i}&=&\frac1{U_{ix}}  \left(\begin{matrix} 1 & -U_{iy} \\{-U_{iy}}
& |U_i|^2\end{matrix}\right), \quad \quad \label{mm}
 \end{eqnarray} (and similarly for $M_{T_i}$)  parameterizes
each $\frac{SL(2)}{U(1)}$ factor.} Each factor of the
$\mathbb{Z}_2\times\mathbb{Z}_2$ orbifold action on the
$SU(4)$ vectors $\widehat{e}^i $
(equivalent to $SO(6)$ spinors) acts according to
\begin{eqnarray}
\mathbb{Z}_2 &
:&(\widehat{e}^1,\widehat{e}^2,\widehat{e}^3,\widehat{e}
^4)\rightarrow(-\widehat{e}^1,-\widehat{e}^2,\widehat{e}^3,\widehat{e}^4)\, ,\\
\mathbb{Z}_2&
:&(\widehat{e}^1,\widehat{e}^2,\widehat{e}^3,\widehat{e}
^4)\rightarrow(\widehat{e}^1,-\widehat{e}^2,-\widehat{e}^3,\widehat{e}^4)\,
,
\end{eqnarray}
so that only the gravitino along the $\widehat{e}^4$ direction remains in
the spectrum.

 Comparison of the truncated algebra (\ref{AlgebraSW}) with the duality
completion of the flux algebra leads to the identifications
\cite{acr}
 \beqa
&&2f_{+abp} = - F'_{abp} \ , \ \ \ \ \ 2f_{+ab}^p =  Q'^p_{ab}  \ ,
\ \ \ \ \ 2f_{+ a}^{bp} = - Q^{bp}_a \ , \ \ \ \ \   2f_{+}^{abp} =
\tilde F^{abp}\, ,
\nn \\
&& 2f_{-abp}= - H'_{abp} \ , \ \ \ \ 2f_{-ab}^p = P'^p_{ab} \ , \ \
\ \ \ \;
 2f_{-a}^{bp} = - P^{bp}_a  \ , \ \ \ \ \  2f_-^{abp} =  \tilde H^{abp}
\, .\label{dic}\eeqa
The factors 2 and signs are  needed to match the normalizations of
(\ref{scag}) and the scalar potential derived from the
superpotential (\ref{sl}).

Let us stress that   comparison of the orbifold projected scalar
potential (\ref{scag}) with the scalar potential derived from the
${\cal N}=1$ superpotential (\ref{sl}) indeed shows
\cite{acr} that both potentials coincide when fluxes and gaugings
are identified as in (\ref{dic}) and truncated identities
(\ref{ji}) are enforced (we will proceed in a similar way below to
obtain ${\cal N}=8$ constraints from comparison of a truncated
${\cal N}=8$ scalar potential with the ${\cal N}=4$ potential
(\ref{scag})). The expressions that we derive below greatly simplify
this comparison.

The relevant  shift matrices acquire a suggestive form
in terms of  the full superpotential (see \ref{sl}) when written
in the $SU(4)$ basis, namely

\begin{equation}
 \begin{array}{ccccccc}
\hspace{1.7cm} A_1^{44} & = &  -\frac{3}{2}ie^{\frac{K}{2}}W &,& \hspace{1.7cm}
A_2^{44} & = & \frac{3}{2}ie^{\frac{K}{2}}W(-\bar S)\, ,\\
\bar A_{2 a4}{}^{ 1}(\bar\gamma^{a})^{ 14} & = &
\frac{i}{2}e^{\frac{K}{2}}W(-\bar U_1) &,& \bar A_{2 a4}{}^{
1}(\bar\gamma^{a})^{ 23} & = &
-\frac{i}{2}e^{\frac{K}{2}}W(-\bar T_1)\, ,\\
\bar A_{2 a4}{}^{ 2}(\bar\gamma^{a})^{ 24} & = &
\frac{i}{2}e^{\frac{K}{2}}W(-\bar U_2)
 &,& \bar A_{2 a4}{}^{ 2}(\bar\gamma^{a})^{ 31} & = &
-\frac{i}{2}e^{\frac{K}{2}}W(-\bar T_2)\, ,\\
\bar A_{2 a4}{}^{ 3}(\bar\gamma^{a})^{ 34} & = &
\frac{i}{2}e^{\frac{K}{2}}W(-\bar U_3) &,& \bar A_{2 a4}{}^{ 3}(\bar
\gamma^{a})^{ 12} & = & -\frac{i}{2}e^{\frac{K}{2}}W(-\bar T_3)\, ,
\end{array}\label{identificacion}
\end{equation}
where $W(-\bar\phi_{i})$ indicates that $\phi_i$ must be replaced by
$-\bar\phi_i$ in the argument of $W$ and all other arguments remain unchanged.
In fact, the superpotential  has a linear
dependence on all the moduli,
which guarantees that
\begin{equation}
(\phi_i+\bar \phi_i) D_{\phi_i}W(\phi_i)=-W(-\bar
\phi_i),
\label{susycondi}
\end{equation}
where  $D_{\phi_i}W(\phi_i)$ is the K\"ahler covariant derivative and
$\phi_i$ stands for $S, T_i$ or $U_i$. The projected shift matrices
can then be written in terms of the superpotential and its K\"ahler
derivatives, as expected.
 Thus, a supersymmetric vacuum solution ensures the
vanishing of the covariant derivatives or, equivalently, the
vanishing of the dilatini ($A_2$) and gaugini ($A_{2a}$) shift
matrices. Such SUSY solution would be Minkowski if $A_1=0$ or Anti
de Sitter otherwise.

\subsection{Spontaneous supersymmetry breaking}

The previous analysis gives us some hints for the study of  spontaneous
 breaking of ${\cal N}=4$
to ${\cal N}=1$ supersymmetry (see  \cite{sw, Dall'Agata:2009gv}).
We must look for a fermionic direction $\epsilon^i$  such that
$\delta \psi_{\mu}^i=0$, $\delta\chi^i=0$ and $\delta\lambda_a^i=0$
in (\ref{fermionvar}) \footnote{ Let us stress that here we do not impose the
orbifold projection so that all the $1+9+9$ complex moduli and all
gaugings $f_{\alpha ABC},  \xi _{\alpha A}$ are in principle
allowed.}. Following \cite{sw}, we define $\epsilon^i= \widehat{e}^i
\xi$ proportional to the $SU(4)$ vector $\widehat{e}^i$, and then
any supersymmetic vacuum configuration must obey
\begin{equation}
 \begin{array}{ccccc}
A_1^{ij}\widehat{e}_j=\sqrt{-\frac{3}{4} V}\widehat{e}^i & ; &
\widehat{e}_jA_2^{ji}=0 & ; & A_{2 aj}^{\quad
i}\widehat{e}^j=0
\end{array}\, . \label{spontSUSYbreak}
\end{equation}
In particular, if we choose $i=4$, any SUSY configuration for
diagonal moduli of the previously discussed orbifold truncation, would vanish
along this direction if the non-diagonal moduli are set to zero and the
JI  (\ref{ji}) are satisfied. Such a configuration is
guaranteed to correspond to a vacuum of ${\cal N} = 4$. Moreover, in
this situation it can be checked that the shift matrices in the
remaining three directions 1, 2, 3 (which we label with indices $i,j,k$), namely
\begin{eqnarray}
A^{kk}_1 &=&-\frac{3}{2} i\ e^{K/2}\ W (-\bar T_j, -\bar U_j)
\ \ \ \ \ \ \ \ \ \ \ \ \ \ \forall j \neq k \, ,\\
A^{kk}_2 &=&\ \ \frac{3}{2} i\ e^{K/2}\ W(-\bar S,-\bar T_j, -\bar
U_j) \ \ \ \ \ \ \ \ \ \forall j \neq k \, ,\\
\bar A_{2aj}^{\ \ \ \ i} (\bar \gamma^a)^{ij}&=&\ \  \frac{i}{2}\
e^{K/2}\ W(-\bar U_i, -\bar T_k)
\ \ \ \ \ \ \ \ \ \ \ \ \ \ \ \  \forall k \neq j \, ,\\
\bar A_{2ai}^{\ \ \ \ 4} (\bar \gamma^a)^{4i}&=&\ \  \frac{i}{2}\
e^{K/2}\ W(-\bar U_j, -\bar T_k) \ \ \ \ \ \ \ \ \ \ \ \ \  \ \ \
\forall j\ , \ \ \ \forall
k \neq i \, ,\\
\bar A_{2ak}^{\ \ \ \ i} (\bar \gamma^a)^{j4}&=&\ \  \frac{i}{2}\
\epsilon_{kij} e^{K/2}\ W(-\bar U_i,\ -\bar T_i)   \ \ \ \ \ \ \ \ \
\ \ \  \forall l
\neq k \, ,\\
\bar A_{2ak}^{\ \ \ \ 4} (\bar \gamma^a)^{ij}&=&\ \  \frac{i}{2}\
\epsilon_{kij} e^{K/2}\ W(-\bar U_m,\ -\bar T_l)   \ \ \ \ \ \ \ \ \
\ \  \forall l\ , \ \ \ \forall m \neq k\, ,
\label{rotesp}\end{eqnarray}
are generically non vanishing and, thus, only one supersymmetry is
generically preserved by the solution. Of course, one should
explicitly check whether these components vanish or not. The former case
would correspond to a solution that spontaneously breaks SUSY to
$1<{\cal N}<4$. We will present some explicit examples of spontaneous
${\cal N} = 4 \to 1$ breaking  below.

\subsection{Including sources}

From a string perspective, part of the constraints (\ref{ji}) correspond
 to
consistency conditions such as Bianchi identities or tadpole
cancellation. In the presence of localized sources, they are
expected to receive  contributions and be non vanishing. So far, the
only localized source that we have considered is the O3 plane. The
orientifold breaks ${\cal N} = 8$ to ${\cal N} = 4$, and (before
taking the orbifold projection) no further breaking is produced if
arbitrary numbers of D3-branes are incorporated in the setup. This
is reflected in the fact that the condition
\begin{equation} \frac{1}{2 \cdot 3!}\ \tilde F^{abc} H_{abc} = N_{D3/O3}
\end{equation}
is not a constraint of ${\cal N} = 4$ supergravity \cite{acr} but
$\tilde F^{abc} H_{abc} = 0$ is a constraint of ${\cal N} = 8$
\cite{deWit:2003} (we will come back to this point in the next section).

Let us recall the construction of the scalar potential
arising in IIB orbifolds when only RR and NSNS 3-form fluxes and
D3-branes are present. This will give us some insight on the role of
branes in the relation between  string compactifications and
truncations of gauged supergravity.

 The IIB
supergravity action in $D = 10$ written in standard notation \cite{Giddings:2001yu} is
\begin{eqnarray}
 S_{IIB}&=&S_{CS}+S_{loc} + \frac{1}{16 G_{10}^2}\int d^{10} x\
 e_{10}\ \left\{e^{-2\phi}\left[R_{10}+4(\partial\phi)^2\right]-\frac{1}{2} F_{(1)}^2 \right. \\
 &&\ \ \ \ \ \ \ \ \ \ \ \ \ \ \ \ \ \ \ \ \ \ \ \  \ \ \ \ \ \ \ \ \ \ \ \ \ \ \ \ \ \ \ \ \ \ \left. -\frac{1}{2\cdotp3!}G_{(3)}\cdotp\overline{G}_{(3)}
-\frac{1}{4\cdotp5!}\tilde{F}_{(5)}^2\right\}\, ,\nn
\end{eqnarray}
where the Chern-Simons and localized sources terms contribute
\begin{eqnarray}
S_{CS} &=&\frac{1}{8iG_{10}^2}\int e^{\phi}\ C_{(4)}\wedge
G_{(3)}\wedge \overline{G}_{(3)}\, ,\\
 S_{loc}&=&-T_3\int d^4x\ e_4\ e^{-\phi}+\mu_3\int_{M^4}C_{(4)} \ , \
 \ \ \  \ \ T_3=\mu_3\, ,
\end{eqnarray}
and the RR 5-form satisfies the self-duality condition
\begin{equation}
\tilde{F}_{(5)}=dC_{(4)}-\frac{1}{2}C_{(2)}\wedge
H_{(3)}+\frac{1}{2}B_{(2)}\wedge F_{(3)} \ , \ \ \ \ \
\tilde{F}_{(5)}=\ast \tilde{F}_{(5)}\, .
\end{equation}
When  D3/O3 sources  are present, the Bianchi identity for
$\tilde{F}_{(5)}$ becomes
\begin{equation}
 F_{(3)}\wedge H_{(3)}=2\ G_{10}^2\ T_3\ \rho_3^{loc} \ \ \ \ \ \ \ \leftrightarrow \ \ \
 \ \ \ \ \frac{1}{2 \cdot 3!}\ \tilde F^{abc} H_{abc} = N_{D3/O3} \label{BianchiF5}\, .
\end{equation}
Replacing  the expectation values of the fields and their respective
fluxes, and taking the internal manifold to be
$T^6/\mathbb{Z}_2\times \mathbb{Z}_2$, one arrives at
\begin{equation}
S_{IIB}=\frac{1}{16 G_{10}^2}\int d^{10}x \ e_{4}\ e_{6}\
\left\{e^{-2\phi}R_{10}-\frac{1}{2\cdotp3!}G_{(3)}\cdotp\overline{G}_{(3)}\right\}
-T_3\int d^4x\ e_4\ e^{-\phi}\, .
\end{equation}
Using the Bianchi identity (\ref{BianchiF5}) and performing the
integration over the internal space leads to
\begin{equation}
 S_{IIB}=\frac{V_6}{16 G_{10}^2}\int d^4x\ e_{4}\ e^{-2\phi}R_{4}-\frac{e^{-\phi}{V_6}}{G_{10}^2}\int d^4x\ e_{4}\ V_{N=1}\, ,
\end{equation}
where $\int e_6=V_6$,
\begin{equation}
V_{N=1} =  \frac{1}{16}f_{\alpha mnp}f_{\beta
qrs}\left[\frac{1}{3}M^{mq }M^{nr}M^{ps}M^{\alpha\beta}
-\frac{4}{9}\epsilon^{\alpha\beta} M^{mnpqrs}\right]\, ,
\end{equation}
and  we have used the dictionary (\ref{dic}).
Notice
that since the only
fluxes present in this example are $F_3$ and $H_3$,
(a more general version of this computation
is presented in \cite{Villadoro:2005cu}  in the IIA framework)
terms involving the metric $\eta$ cannot appear in the potential because
they necessarily contain non geometric fluxes. Therefore, this
proves that this string compactification in the presence of an
arbitrary number of D3-branes gives rise to a potential with the structure of
a truncation of an ${\cal N} = 4$ gauged supergravity, provided the
Bianchi identity (\ref{BianchiF5}) is used. We emphasize that in
this case, for any D3/O3 charge, it preserves ${\cal N} =
4$, so this example corresponds to an exact truncation of ${\cal N} =
4$ gauged supergravity.

If SUSY-breaking sources\footnote{By SUSY-breaking sources we refer
to BPS branes, which break half of the supersymmetries.} (like
D7-branes) were introduced together with O3-planes, their tadpole
cancelation conditions would violate the ${\cal N} = 4$ constraints,
and the scalar potential would no longer correspond to an ${\cal N}
= 4$ truncation, but rather to a deformation. In fact, these objects
project out a different set of gravitini than those projected by the
O3-plane, and therefore,  including them necessarily breaks
supersymmetry partially. This is reflected in the fact that the
consistency conditions of ${\cal N} = 4$ supergravity (\ref{ji})
force the D7-brane charges (and also their S-duals) to
vanish\footnote{Recall that we are restricting to the case $n = 0$.
For $n>0$, gauged ${\cal N}=4$ supergravity admits effective $(p,q)7$-brane
charges when no other SUSY-breaking sources are present
\cite{acr}.}. Similarly, the constraints (\ref{ji}) can also
be sourced by other SUSY-breaking objects, such as KK-monopoles
\cite{Villadorosource, Villadoro:2007yq} or other dual (exotic)
objects.

In the presence of such SUSY-breaking sources, the structure of the
(super)potential  in terms of gaugings and scalars is formally the
same as that of a supergravity truncation, but in both cases the
parameters satisfy different constraints: on the string side, tadpole
conditions allow the inclusion of (SUSY-breaking) brane charges, for
example for seven branes
\begin{equation} N_{D7i} = (QF)_i \ , \ \ \ \ \ N_{I7i} = (QH + PF)_i\ , \ \ \
\ \ N_{NS7i} = (PH)_i \, ,\end{equation}
while gauged ${\cal N} = 4$ supergravity requires that these charges
vanish, i.e. $N_{D7i} = N_{I7i} = N_{NS7i} = 0$. Therefore, in the
presence of SUSY breaking sources, the untwisted sectors of string
compactifications are not truncations of gauged supergravities. They
are discrete {\it deformations} of truncations, and only at some
non-generic points of their parameter space (when there are no
sources) they are exact truncations.

\subsection{New constraints from gauged ${\cal N}=8$ supergravity truncations}
Our aim here is to explore flux compactifications with an underlying $ {\cal
N}=8$ supergravity theory. Namely, ${\cal N}=1$ compactifications where
the untwisted sector is a projection of an  $ {\cal N}=8$  supergravity
theory.
In particular, we are interested in the derivation
of possible  new constraints on dual fluxes.

Recall from the previous section that extra constraints are
expected. Specifically, no brane (or orientifold)  charge should be
allowed at all if $ {\cal N}=8$ is to be preserved. Therefore,
besides the ${\cal N}=4$ constraints, we also expect to obtain
\begin{equation} \frac{1}{2 \cdot 3!}\ \tilde F^{abc} H_{abc} = N_{D3/O3} = 0
\, ,
\label{FHD3}\end{equation}
ensuring that no  D3/O3 charges are present.

A possible strategy to read the constraints on fluxes is to compare
the respective scalar potentials. Namely, we must identify the
conditions to impose on the ${\cal N}=4$ scalar  potential
so that it coincides with
the truncated $ {\cal N}=8$ one. From
\cite{samt,Weidner:2006rp} we find
\begin{equation}
V_{\mathcal{N}=8}=  {\cal X}^{\cal L}_{\cal A \cal B} {\cal X}^{\cal
S}_{\cal P \cal Q} M^{\cal A \cal P}M^{\cal B \cal Q}M_{\cal L \cal
S}+7{\cal X}^{\cal Q}_{\cal A \cal B} {\cal X}^{\cal B}_{\cal P \cal
Q} M^{\cal A\cal P}\, , \label{n8sp}
\end{equation}
where calligraphic indices ${\cal A}, {\cal B},...$ span the $\bf
{56}$ vector representation of the ${\cal N}=8$ duality group $E_7$
(we refer to \cite{samt,Weidner:2006rp} for details).
 The fundamental
$\rep{56}$ representation of $\E7$ decomposes under $\Tsub \subset \E7$ as
\begin{equation}
\label{56}
\begin{aligned}
   \rep{56} & = (\rep{12},\rep{2}) + ({\rep{32'}},\rep{1})\ ,\\
   \lambda & = \left( \lambda^{A\alpha},\, \lambda^{s-} \right)\, ,
\end{aligned}
\end{equation}
where (greek labels)  $\alpha=1,2$ and $A=1,\dots ,12$,
are the indices that will
survive the
projection to  ${\cal N}=4$. Thus, in terms of this decomposition, we can write
the metric matrix as
\begin{eqnarray}
M^{\cal A \cal P}&=& \left(\begin{matrix} M^{ A\alpha,\,\beta P} & M^{
A\alpha,\,
s-}\\ M^{s-,\,  A \alpha}
& M^{s-,\, s-}\end{matrix}\right), \quad \quad
 \end{eqnarray}
while the contributions from the gaugings that survive the projection
are \cite{sw,aacg}
\begin{equation}
{\cal X}^{\cal P}_{\cal A \cal B}= {\cal X}^{ P\rho }_{A\alpha
B\beta }= \delta^\rho _\beta  f_{\alpha AB}^P -
\frac{1}{2}\left(\delta^P_A\delta^\rho_\beta \xi_{\alpha
B}-\delta^P_B\delta^\rho _\alpha\xi_{\beta  A} - \delta^\rho _\beta
\eta_{AB}\xi^P_{\alpha} + \epsilon_{\alpha \beta }\delta^P_B
\xi_{\delta A}\epsilon^{\delta \rho }\right)\, .
 \end{equation}
Replacing them into (\ref{n8sp}),
 we find (the $\xi_{\delta
A}$ terms are not written, for simplicity)
\begin{eqnarray}
 V_{\mathcal{N}=8}&=& \delta_\beta ^l f_{iAB}^L
 \delta_\kappa^s f_{\rho PQ}^S
M^{ A\alpha \, P\rho}M^{ B\beta   Q\kappa}M_{ L\lambda  S\sigma }+7
\delta_\beta ^\kappa f_{\alpha AB}^Q
 \delta_\kappa^\beta  f_{\rho PQ}^B
M^{ A\alpha P\rho}+\dots\nonumber\\
&=& \delta_\beta ^\lambda  f_{\alpha AB}^L
 \delta_\kappa^\sigma  f_{\rho PQ}^S
M^{ A\alpha \, P\rho}M^{ B\beta   Q\kappa }M_{ L\lambda  S\sigma }+14
 f_{\alpha AB}^Q
 f_{\rho PQ}^B M^{ A\alpha P \rho}+\dots \ ,
\end{eqnarray}
where the dots indicate  contributions involving indices in the
$({\rep{32'}},\rep{1})$ spinorial representation that are projected out in the
truncation.

In the appropriate basis, the matrix $M^{ A\alpha P\rho}$ can be
written as a product of factors $ M^{ \alpha \rho} M^{ A P}$, with
$M^{ \alpha \rho}$ and $ M^{ A P}$  the metric matrices defined in
(\ref{msu2}) and (\ref{md6}), respectively. The factors depend on
the explicit choice of basis. This is similar to what is explained
in footnote 3.

We finally obtain
\begin{eqnarray}
 V_{\mathcal{N}=8}&=& \delta_\beta ^\lambda  f_{\alpha AB}^L
 \delta_\kappa^\sigma  f_{\rho PQ}^S
M^{ \alpha \, \rho}M^{ A \, P}
M^{\beta  \kappa}M^{ B  Q}
M_{ \lambda  \sigma }M_{ L S}
+14f_{\alpha AB}^Qf_{\rho PQ}^B M^{ \alpha \rho }M^{ A P}+\dots\nonumber\\
&=&  f_{\alpha AB}^L
  f_{\rho PQ}^S
M^{ \alpha \, \rho }M^{ A \, P}
M^{ \lambda   \sigma }M^{ B  Q}
M_{ \lambda  \sigma }M_{ L S}
+14f_{\alpha AB}^Qf_{\rho PQ}^B M^{ \alpha \rho }M^{ A P}+\dots\nonumber\\
&=& 2 f_{\alpha AB}^L
  f_{\rho PQ}^S
M^{ \alpha \, \rho }M^{ A \, P} M^{ B  Q} M_{ L S} +14f_{\alpha
AB}^Qf_{\rho PQ}^B M^{ \alpha \rho }M^{ A P}+\dots\, ,
\end{eqnarray}
where we have used that $M^{ \lambda   \sigma }M_{ \lambda  \sigma
}=\delta_\lambda ^\lambda =2$.

Thus, besides some normalization factors, when compared with the
${\cal N}=4$ scalar potential,
 we
find that in order for both potentials to match, the following set
of quadratic constraints must be satisfied
\begin{eqnarray}
&&f_{\alpha ABC}f_{\beta }^ {ABC}=0\, ,\label{Singlet}\\
&&\epsilon^{\alpha\beta }f_{ \alpha[ABC}f_{\beta  PQR]}=0 \,
.\label{FHCov}
\end{eqnarray}
Interestingly enough,  equations (\ref{FHCov}) are the covariant
generalization of the tadpole cancelation condition for D3-branes
in the absence of sources, i.e. (\ref{FHD3}). As explained, this
constraint in particular is expected because such localized sources
necessarily break maximal supersymmetry. One can check explicitly
through a systematic application of $SL(2,\mathbb{Z})^7$
transformations over (\ref{FHD3}) that the full set of constraints
(\ref{FHCov}) is obtained in the orbifolded case. Formally, one could
also obtain these constraints from the JI satisfied by gauge
generators in the $E_7$ invariant theory and further projecting to
 $\Tsub \subset \E7$. Steps are indicated in \cite{aacg}.

Looking carefully at the scalar potential (\ref{scag}), it can be
checked that the terms that cannot be embedded in ${\cal N} = 8$ are
precisely those that incorporate tree-level mass terms for the
scalars (the remaining terms being interactions). This might
find its origin in the fact that gauged ${\cal N} = 8$ supergravity
cannot contain tree-level mass terms for scalars because these
belong to the same supermultiplet as the graviton. When
supersymmetry is broken, the graviton multiplet remains massless, but
the scalars belonging to the vector multiplets of ${\cal N} = 4$ can
develop a mass, and therefore the vanishing mass conditions
(\ref{Singlet})-(\ref{FHCov}) are not requirements of gauged ${\cal
N} = 4$ supergravity.

\section{Analysis of moduli stabilization }\label{sec4}

In this section we explore vacuum solutions of untwisted sectors of the
Type IIB string compactifications discussed above. We begin by recalling
generalities related to moduli fixing, mainly focusing on the
constraints that fluxes and solutions must satisfy. We observe that
if the only sources allowed to be present in the configuration are
D3-branes and $(p,q)7$-branes, the constraints favour flux
configurations in which magnetic and electric gaugings are
proportional to each other, and we show that in such case complete
moduli fixing cannot be achieved. This allows us to exclude large
regions of the parameter space, and concentrate on models with
potentiality to fully stabilize all moduli in string orbifold
compactifications. We conclude the section with the analysis of some
representative examples.
\subsection{Generalities}
\subsubsection{Constraints on fluxes and solutions}

Consistent searches for
 minima of the scalar potential
require that  the fluxes and moduli satisfy several physical and
algebraic constraints. The former include the following list:
\begin{enumerate}
\item The stabilized moduli must satisfy
\beqa
{\rm Re}~ T_i  &=& t_x=e^{-\phi} R_j R_{j+3} R_k R_{k+3}\ >\ 0 \, ,\\
{\rm Re}~ U_i  &=& u_x=\frac{R_{i+3}}{R_i}\ >\ 0 \, ,\eeqa $R_i$
being the radii of the factorized tori in the absence of fluxes.
Strictly speaking, when introducing fluxes the internal space warps
and these expressions are expected to change. We assume that in a
large volume scenario, in which the fluxes are diluted, these
constraints hold.

\item The vacuum must be stable (for SUSY configurations this is
automatically verified, see \cite{Borghese:2010ei} for SUSY-breaking
vacua).

\item Solutions to the equations of motion must fall in the perturbative regime
(${\rm Re}~ S = e^{-\phi} \gg 1$) and the volume of the internal
space vol$_6 = e^{3 \phi / 2} ({\rm Re} T_1\ {\rm Re} T_2\ {\rm Re}
T_3)^{1/2}$ should be large, in order for this scenario to be self
consistent and not sensitive to higher order corrections.

\item Other possible physical constraints determined by
phenomenology. In particular, this includes the requirement that all
scalars must be massive in the vacuum.

\end{enumerate}

In addition, we know from the previous section that many consistency
constraints must be obeyed by the fluxes, generically taking the
form
\beqa f_{\alpha P}^{[AB} f_{\beta}^{CD] P}&=& S^{ABCD}_{\alpha\beta}\, ,\\
\epsilon^{\alpha \beta} f_{\alpha P}^{AB} f_{\beta}^{CD P}  &=&
K^{ABCD}
\, ,\\ \epsilon^{\alpha
\beta} f_\alpha^{[ABC} f_\beta^{DEF]} &=& T^{ABCDEF} \, , \label{FHCov2} \\
f_\alpha^{ABC} f_{\beta ABC} &=& R_{\alpha \beta}\, .
\label{contraidos}\eeqa
We have also discussed in the previous section that embeddings into
${\cal N} = 4$ gauged sugra require
\begin{equation} S^{ABCD}_{\alpha\beta} = 0 \ , \ \ \ \ \ K^{ABCD} = 0
\, ,\end{equation}
and further embeddings into ${\cal N} = 8$ gauged sugra must verify
\begin{equation} T^{ABCDEF} = 0 \ , \ \ \ \ \ R_{\alpha \beta} = 0
\, .\end{equation}
Many components of these tensors were identified with brane charges.
For example, the O3/D3  charge is given by $T^{123456}$, and
$(p,q)7$-brane charges wrapping the 4-cycles $abcd$ are
parameterized by $S_{\alpha \beta}^{abcd}$, the sub-indices
$\alpha,\beta$ labeling the different types of $(p,q)7$-branes (see
 \cite{acr} for details).

In this section we consider deformations of truncated
${\cal N} = 4, 8$  gauged  supergravities in the presence of D3-branes and
$(p,q)7$-branes. By deformed ${\cal N} = 4$ we mean
\begin{equation} S^{ABCD}_{\alpha\beta} = \left\{\begin{matrix}S_{\alpha\beta}^{ABCD} \ \ \ \ \ \  \ \ A,B,C,D \leq 6 \\   0 \ \ \ \ \ \ \ \ \ \ \ \ \ \ \ {\rm otherwise}\ \ \ \ \ \ \end{matrix}\right. \ , \ \ \ \ \ K^{ABCD} = 0\end{equation}
and  ${\cal N} = 8$ deformations additionally satisfy
\begin{equation} T^{ABCDEF} = \left\{\begin{matrix}4\cdot 3! \cdot N_{D3/O3} \ \ \ \ \ \  \ \ ABCDEF = 123456 \\ \ \ \  0 \ \ \ \ \ \ \ \ \ \ \ \ \ \ \ \ \ \ \ \ \ \ {\rm otherwise}\ \ \ \ \ \ \ \ \ \ \ \ \ \ \ \end{matrix}\right. \ , \ \ \ \ \ R_{\alpha \beta} = 0\, .\end{equation}
Of course, other deformations are possible\footnote{For example, in
\cite{Derendinger:2004jn} some deformations involving $K^{ABCD} \neq
0$   were considered in the IIA picture. Such deformations are sourced by
KK-monopoles
\cite{Villadorosource, Villadoro:2007yq, Dall'Agata:2009gv}.}
 but we will not consider
them here.

~

To solve the equations for fluxes we  pursue the following strategy.
We  start with the following subset of equations mixing the $f_+$
(electric) and the $f_-$ (magnetic) sectors: \beqa \epsilon^{\alpha
\beta} f_{\alpha P}^{AB} f_{\beta}^{ CD P}  &=& 0 \ \ \
\ \ \forall\ A,B,C,D \ \ \ \ \ \ \ \ \ \ \ \ \ \ \ \ \ \ ({\rm for \ deformed}  \ {\cal N}= 4,8)\, , \label{N48}\\
\epsilon^{\alpha \beta}f_\alpha^{[ABC} f_\beta^{DEF]} &=& 0\ \ \ \ \
\forall\ ABCDEF \neq 123456\ \ \ \ ({\rm for \ deformed }  \ {\cal
N}= 8) \, ,\label{N8}\eeqa and consider the electric
 and  magnetic gaugings respectively as parameters and variables
 (they can obviously
be interchanged) of a linear system of equations schematically
taking the form \beq {\cal F}_{ij}(f_+) f_-^j = 0 \ , \label{Linear}
\eeq
where $j$ is summed over the 64 magnetic fluxes
$f_-^j$ and $i$ labels equations (\ref{N48}) and (\ref{N8}). The entries of the
matrix ${\cal F}_{ij}$ are given by the values of the electric fluxes and are
univocally determined by equations (\ref{N48}) and (\ref{N8}).
A given set of non-zero  $f_+$ fluxes defines the linear system
(\ref{Linear}), which can then be linearly solved for $f_-$.
Finally, the solution is replaced in the remaining equations,  which
in general happen to be easily solvable. See \cite{deRoo:1985jh,
deCarlos:2009qm} for interesting group-theoretical approaches
to solve some of these equations.

We would like to point out two general difficulties one encounters
when looking for vacua. First, as can be seen in the superpotential
(\ref{sl}), when all the fields are rescaled as $\tilde
\phi = i \phi$, one ends with a superpotential with real-valued
coefficients for $\tilde \phi$. Therefore, it is only possible to
stabilize all $\tilde \phi$ at complex values when the models have
enough structure, a situation difficult to achieve due to the
enormous amount of algebraic constraints on the fluxes. Second,
since (\ref{N48})-(\ref{N8}) are antisymmetric ($\epsilon^{\alpha
\beta} f_\alpha f_\beta = 0)$,  most
 of the solutions are of the form $f_+ = \alpha f_-$. We
will show in the next subsection that in such cases it is impossible
to stabilize all the moduli, and thus it will allow us to exclude large
regions of the parameter space.

\subsubsection{Exclusion argument for stabilized vacua} \label{sectionArg}
In this subsection we present a general result that will have some
consequences for moduli fixing.  The superpotential
(\ref{superpotgeneral}) and the K\"ahler potential (\ref{PotKahler})
have the following structure \beq W = (f_{+I} -i S f_{-I}) M^I(T,U)
\ , \ \ \ \ \ \ K = -\log(S+\bar S) + \tilde K(T,U)\, ,
\label{superkahlergen} \eeq  and lead to a scalar potential of the
form \beq V = (M^{\alpha \beta} + i \epsilon^{\alpha \beta})
f_{\alpha I} f_{\beta J} {\cal F}^{IJ}\, , \label{potescalargeneral}
\eeq where $\epsilon^{+-} = 1$, $ M^{\alpha\beta}$ is given in
(\ref{mm}) and \beq {\cal F}^{IJ} = \frac{1}{2} e^{\tilde K}
\left[(\partial_A M^I + M^I \partial_A \tilde K) \tilde K^{A \bar B}
(\partial_{\bar B} \bar M^J + \bar M^J\partial_{\bar B}\tilde K ) -
3 M^I\bar M^J + \bar M^IM^J\right] \, .\eeq

As argued in the previous subsection, the gauged supergravity
constraints are solved in many cases by configurations in which
electric and magnetic gaugings are proportional, $i.e.$ $f_{+I} =
\alpha f_{-I}$ with $\alpha \in \mathbb{R}$. In such cases, the
scalar potential (\ref{potescalargeneral}) becomes \beq V =
\frac{1}{s_x} \left((\alpha + s_y)^2 + s_x^2 \right) \Omega \ , \ \
\ \ \Omega = f_{- I} f_{- J} {\cal F}^{IJ}\, . \eeq When looking for
fixed points, one must focus on its first derivatives, namely \beqa
\frac{2}{ s_x} (\alpha + s_y)\ \Omega
&=& 0\, ,\label{derImS}\\
-\frac{1}{ s_x^2}\left((\alpha + s_y)^2 - s_x^2\right) \Omega
&=& 0 \, ,\label{derReS}\\
\frac{1}{ s_x}\left((\alpha + s_y)^2 + s_x^2\right) \partial_i
\Omega  &=& 0\, . \label{ultimaEOM}\eeqa As can be seen,  $S$ can
only be fixed  at $S = - i \alpha$, which
corresponds to a singular value of the equations. This analysis,
based on the axion-dilaton modulus $S$, can be extended to the other
moduli. In fact, using the spinor formalism, one can write the
superpotential and scalar potential in a more {\em democratic} way as
\beqa W &=& i\
p(a^0,\dots , a^6)\ \Phi^{a^0,0}\dots \Phi^{a^6, 6}\nn \\
V &=& p(a^0, ..., a^6) p(b^0, ..., b^6)
M_0^{a^0 b^0} ... M_6^{a^6 b^6}; \  \ \
M_j^{a^jb^j} = \frac{1}{{\rm Re} \phi_j} \left(
\begin{matrix} 1 & {\rm Im} \phi_j
\\ {\rm Im} \phi_j & |\phi_j|^2\end{matrix}\right)\,  , \nn   \eeqa
where $a^j = \pm$, $\Phi^{+,j} = i$ and $\Phi^{-,j} =
\phi_j$ with
$\boldsymbol \phi = (S, \bf T, \bf U)$ and $j=0,...,6$.
The notation for parameters
explicitly indicates their corresponding weight \cite{acr}. Written
in this form, it is easy to see
 that
 $S$ is equivalent to any other modulus, and one can
therefore conclude that it is not possible to stabilize all moduli
when \beq p({a^0},... ,+ ,... ,a^6) = \alpha_j\ p({a^0}, ...,  -
,..., a^6)\, , \label{cond} \eeq
 with the $+$ and $-$ signs at the same
 position $j$. Pairs of dual fluxes must be present and
independent  for moduli fixing.

~

Moreover, this statement can be further extended for Minkowski
vacua. When (\ref{cond}) holds but
 there is additionally a unique non vanishing
$p({a^0}, ...,  \mp ,..., a^6)$ whose dual $p({a^0}, ...,  \pm
,..., a^6)$ is zero, then the additional equation $W = 0$ forces
$p({a^0}, ...,  \mp ,..., a^6)$ to vanish, thus leading to the
previous case. Notice that if the scalar potential is independent of
a field $\phi_i$, since all fields appear linearly in the
superpotential, in any SUSY vacuum one would have a vanishing
 expresion (\ref{susycondi}) and therefore that vacuum
is necessarily Minkowski. This is consistent with the fact that when
electric and magnetic gaugings are proportional, the scalar
potential takes the form
\begin{equation} V = \frac{\alpha + s_y}{s_x} \left[\frac{2}{s_x} (\alpha +
s_y) \Omega\right] - \frac{1}{s_x^2} ((\alpha + s_y)^2 -
s_x^2)\Omega \end{equation} and it can be seen from the equations of
motion (\ref{derImS})-(\ref{ultimaEOM}) that any vacuum is
necessarily Minkowski, i.e. $V = 0$.

\subsubsection{Excluding regions of the parameter space}
Here we apply the results of the previous subsection
 to identify regions of the
parameter space that are excluded for moduli fixing (i.e. the region
in which electric and magnetic fluxes are proportional). This
argument permits to understand why most of the solutions to the
constraints on fluxes are unlikely to fully stabilize moduli.

We restrict to a particular compactification scenario with an
additional $\mathbb{Z}_3$ symmetry, mainly because the parameter
space is smaller than in the general case and strong conclusions can
be obtained. Compactifications with $\mathbb{Z}_3$ symmetry are
probably the most explored setups for moduli fixing. The conclusions
reached in this particular case are representative of the general
setup. The prescription we use is that of \cite{stw}, namely \beq
(T^2)^3 = T_{(1)}^2 \times T_{(2)}^2 \times T_{(3)}^2   \  \ \ \ / \
\ \ \ \mathbb{Z}_3:\ T_{(1)}^2 \to T_{(2)}^2 \to T_{(3)}^2\to
T_{(1)}^2 \, .\eeq This identifies some of the moduli $T_i \to T$
and $U_i \to U$, and also some parameters (see Appendix), thus
reducing the parameter space and making it simpler to look for
solutions. Also, since it rotates the two-tori, the tadpole
cancelation conditions for seven branes wrapping the three different
four-cycles become equal.

We consider separately ${\cal N} = 4$ deformations (involving the
constraints (\ref{N48})) and ${\cal N} = 8$ deformations (involving
(\ref{N48}) and (\ref{N8})). The number of equations contained in
(\ref{Linear}) is much larger in ${\cal N} = 8$ than in ${\cal N} =
4$, so generically the rank of the  $\cal F$ matrix (\ref{Linear})
will be bigger. In either case, when the following relation is
satisfied
\beq {\rm rank} {\cal F} = \# {\rm magnetic \ fluxes}
- 1\, ,
\label{excluded}\eeq
the linear system of equations admits a unique non-trivial solution,
namely $f_+ = \alpha f_-$, and the model falls in the exclusion
region. In this case, where there is a ${\mathbb Z}_3$ symmetry,
$\# {\rm magnetic \ fluxes} =24$
(if this symmetry is removed $\# {\rm magnetic \ fluxes} =64$).
Then, we must look for models satisfying
\beq {\rm rank} {\cal F} < \# {\rm magnetic \ fluxes} - 1\, .
\label{notexcluded}\eeq

\begin{figure}[h]
\includegraphics[width=15 cm]{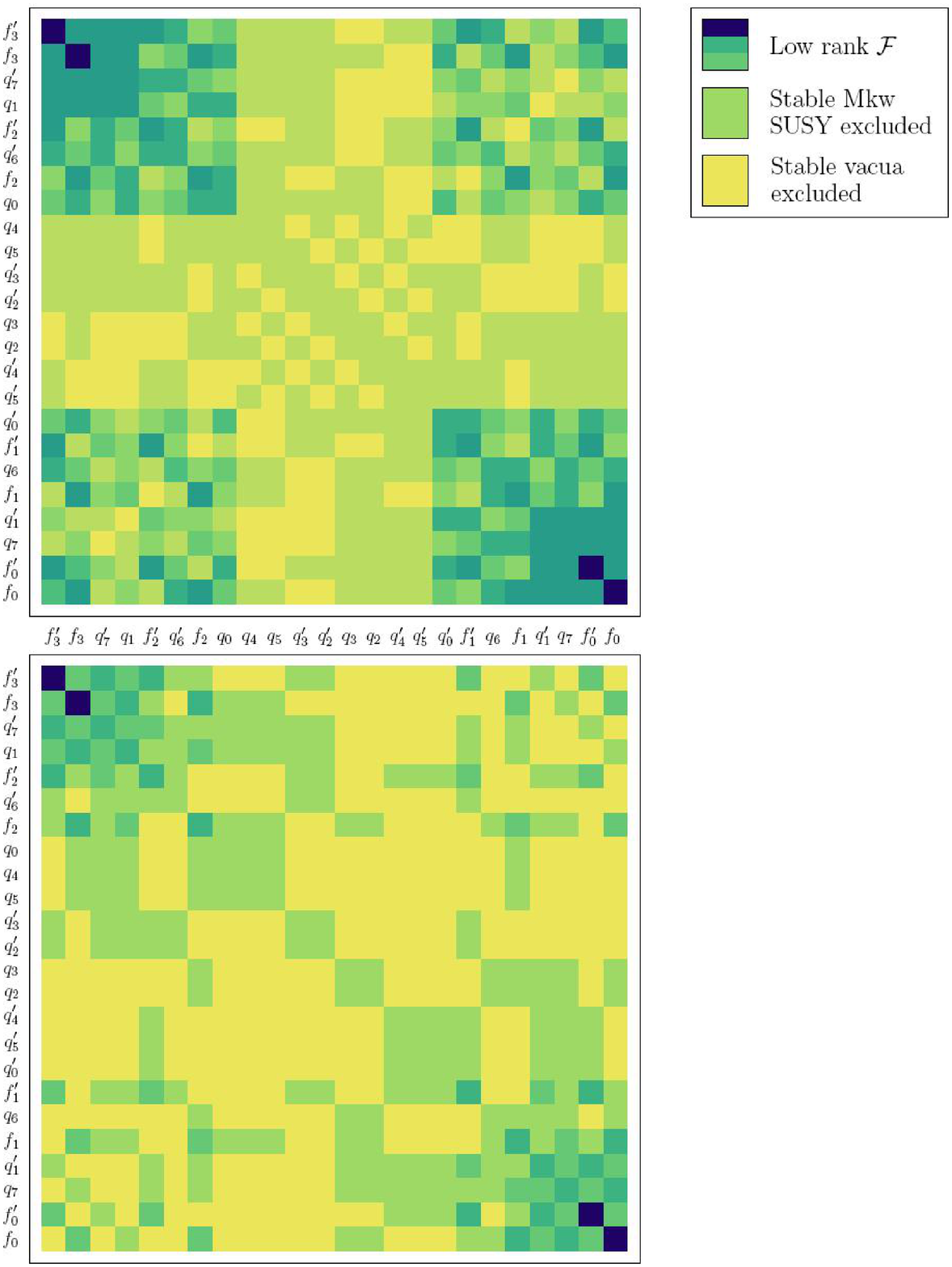}
\caption{The rank of $\cal F$ is represented when one or two generic
electric gaugings are turned on in the ${\cal N} = 4$ (upper graph)
and ${\cal N}= 8$ (lower graph) constraints. Yellow points are
excluded for moduli stabilization because ${\rm rank} {\cal F}= \#
{\rm magnetic \ fluxes} - 1 = 23$.} \label{Figura}
\end{figure}

\begin{itemize}
\item ${\cal N} = 4$ deformations.
In figure \ref{Figura} (upper graph) we sketch rank${\cal F}$ for
all possible sets of one or two $f_+$ generic
parameters\footnote{Non-generic values of the parameters can reduce
the rank of $\cal F$. Here we restric to a generic analysis. }. When
searching for solutions different from $f_+ = \alpha f_-$, the only
interesting cases are those in which (\ref{notexcluded}) is
satisfied. Already at this stage of the procedure, when only two
magnetic parameters are turned on, a large part of the parameter
space is discarded for moduli stabilization in any vacuum. When
adding more non vanishing parameters the situation gets worse
because the rank of $\cal F$ increases and gets closer to the bound
(\ref{excluded}). An explicit counting shows that more than $95\%$
of the parameter space falls in the exclusion region. For Minkowski
vacua, the exclusion region is even larger because already when $
{\rm rank} {\cal F} = \# {\rm magnetic \ fluxes} - 2, $ the models
are not viable for moduli fixing.

\item ${\cal N} = 8$ deformations. For this case, we sketch rank$\cal F$ in figure \ref{Figura}
(lower graph). Notice that the excluded region (when only two
electric gaugings with generic values are turned on) is even much
larger than in the previous case. We have not found models surviving
the exclusion argument that admit fully stabilized vacua with
physically acceptable values for the moduli. This might find its
origin in the fact that the parent ${\cal N} = 8$ theory is massless
and all fields are moduli at tree-level. This should be considered
as a limit, the limit of maximal constraints for these kind of
models.
\end{itemize}

In summary, we have shown why it is so difficult to stabilize vacua
when the constraints are imposed. The extreme  limit is that in
which all ${\cal N} = 8$ constraints are considered in the absence
of sources. When sources are added and the number of
supersymmetries reduces, the possibilities for stabilizing vacua
increase. In the next section we show some representative examples
of SUSY vacua, in different situations.

\subsection{SUSY vacua}

Here we present different examples to illustrate the general results
about moduli fixing, and the possibility to embed string vacua in
(deformed) gauged ${\cal N} = 4,8$ supergravity truncations.

We start with an analysis of Type IIB AdS vacua, restricting to the
case in which only fluxes that are dual to
IIA geometric, RR and NSNS fluxes are turned on. We first
consider a well known example \cite{af} arising from a controlled and
consistent $D = 10$ setup. Then we move to the classification of all
allowed isotropic models with IIA-geometric fluxes satisfying the
${\cal N}=4, 8$ constraints with branes, and exhibit some of the
simplest solutions. Models
involving
$P$-fluxes and admitting fully stabilized vacua
are also displayed.
We also show how  examples that can be embedded
in gauged supergravity  can be
seen as vacua spontaneously breaking ${\cal N} = 4 \to 1$ and  we discuss
further constraints imposed by Freed-Witten (FW) anomalies
\cite{Freed:1999vc}.

We then consider Minkowski vacua. We argue that when primed fluxes
are turned off, the algebraic constraints generically do not allow
stable
 Minkowski vacua satisfying the physical constraints. Finally,
we display some examples.

\subsubsection{Anti de Sitter}

\noindent {\bf The IIB dual of the AdS solution of \cite{af}}

Let us reconsider the model introduced in \cite{af}, T-dualized to
the IIB picture. Interest in analyzing this case
 resides in the fact that it can be uplifted to $D = 10$ as a consistent supergravity
solution in the IIA picture. Additionally, it is known to stabilize
some of the moduli and allows for vanishing brane charges (and
therefore  exact embeddings in gauged supergravity).

The model is defined by the  superpotential \beq W_{AF} = e_0 + i
e_i U_i - q_r U_s U_t + i m U_1 U_2 U_3 + i h_0 S - a_i U_i S - i
h_i T_i - b_{ij} T_j U_i \, ,\eeq
where the following
identification is chosen \beqa
 q_i = - c \ ,  \ \ \ \ \ e_i = e\ ,  \ \ \ \ \ \ \ a_i = a \, ,\ \ \ \ \ \ \
-b_{ii} = b_{ji} =
 b_i \, .
\eeqa

The tadpole cancelation conditions  required by string theory are
given by
 \beq N_{D7i} = \frac{1}{2}\left(m h_i
+ b_i c\right)  \ , \ \ \ \ \ N_{D3/O3} = \frac{1}{2}\left(h_0m -
3ac\right) \label{D7vanishing} \eeq
and all other constraints are automatically satisfied. The first
 constraints define the charge of D7-branes wrapping the
4-cycles $\tilde \omega_i$. If such branes were present $N_{D7i}
\neq 0$, they would select a different projection from that already
determined by the orientifold, and therefore, even if the orbifold
were removed, SUSY would be broken to ${\cal N} < 4$. This is the
physical reason why $N_{D7i} = 0$ is the only constraint to be
imposed in order to match this model with an exact ${\cal N}= 4$
truncation. Here we allow for non-vanishing $N_{D7i}$ and
$N_{D3/O3}$.

This effective theory admits AdS minima provided the fluxes verify
\beq 3 h_k a + h_0 b_k = 0\, . \eeq The vacuum is isotropic in $U_i
= U = u_x + i u_y$. The dilaton and the real part of the K\"ahler
fields are given purely in terms of the complex structure moduli
\beq s_x = \frac{2 u_x (c - m u_y)}{a}\ , \ \ \ \ \ \ \ t_{ix} =
\frac{3 a s_x}{b_i}\, . \eeq The corresponding axions, on the other
hand, are not entirely
fixed\footnote{As discussed in \cite{cfi} in the context of
 Type IIA with D6-branes, massless axions are
necessary in some specific models
for certain (potentially anomalous) brane U(1) fields to get a
St\"uckelberg mass.}, and determine  massless directions
given by \beqa s_y(t_y,u_y) &=& \frac{3e a + 3c (3h_0 - 7 a u_y) - 3
m u_y (3h_0 - 8 a u_y)
- 3 b t_y a }{3 a^2}\, ,\\
t_y &=&\frac{b_1 t_{1y}}{b}\ =\ \frac{b_2 t_{2y}}{b}\ = \ \frac{b_3
t_{3y}}{b}\, . \eeqa Now only the complex structure moduli remain to
be stabilized. Two situations can be distinguished
\begin{itemize}
\item For $m = 0$, the solutions are given by
\beq u_y = \frac{h_0}{3a} \ , \ \ \ \ \ \ u_x^2 =
\frac{1}{9c}\left(e_0 - \frac{h_0 e}{a}- \frac{h_0^2 c
}{3a^2}\right)\, . \eeq

\item For $m \neq 0$, one first has to solve
\beqa &&
3 h_0^2  m (c - m u_y) - a^2 (e_0 m + u_y (10 m u_y - 9 c)(16 m
u_y - 15 c)) \nonumber\\
&& \  \ \ \ \ \ \ \ + ~a h_0(45 c^2 - 106 c m u_y - m (e + 62 m u_y^2)) = 0
\eeqa and replace the positive solutions into \beq u_x^2 = \frac{5  (h_0
- 3 a u_y) (m u_y - c)}{am} \, . \eeq
\end{itemize}

We now want to see if this model can be considered as a deformation of a
gauged ${\cal N} = 4, \ 8$ supergravity truncation. The parameters
of this model automatically satisfy the constraints of the deformed
${\cal N} = 4$ supergravity, and can be seen as an exact truncation
of ${\cal N} = 4$ supergravity if $N_{D7i} = 0$, as we already
explained. Concerning ${\cal N} = 8$, even if we set $N_{D3/O3} =
0$, there is an additional equation that is not satisfied by this
solution, namely
\begin{equation} ab_i =0 \, .\label{N8raro}\end{equation}
One would have thought that in the absence of D3/O3 charges this
theory could be regarded as a truncation of an ${\cal N} = 8$ gauged
supergravity. However surprisingly, the constraint (\ref{N8raro})
cannot be satisfied by the solution.

We would now like to follow the steps of \cite{Dall'Agata:2009gv} in
order to conceive this solution as a spontaneous SUSY breaking
${\cal N} = 4 \to 1$ configuration. To achieve this, one first has
to set $N_{D7i} =0$. Then, as explained, instead of truncating the
non-diagonal moduli through orbifolding as described in Section
\ref{sec0}, one considers these moduli as dynamical fields, but
provides them with vanishing vacuum expectation values. As explained
in Section \ref{sec1}, this leads to a situation in which the ${\cal
N}=4$ shift matrices are exactly equal to the orbifolded ones, but
in this case all the SUSY directions have to be analyzed. Since the
above is an AdS solution of the truncated theory, it will give rise
to a non-vanishing $A_1^{44}$ direction and vanishing $A_2^{44}$ and
$A_{2a4}^{\ \ \ j}$, and therefore it corresponds to a solution of
the parent ${\cal N} = 4$ supergravity if $\epsilon^1 = \epsilon^2 =
\epsilon^3 = 0$ and $\epsilon^4 \neq 0$ are taken. Whether it breaks
supersymmetry or not has to be determined by evaluating the remaning
directions (\ref{rotesp}) of the shift matrices. In particular, we
focus on $A_2$, for which  (when $m = 0$) \beq -i
\frac{2}{3}e^{-K/2}A_{2}^{kk} \propto \frac{a^2\sqrt{b_1 b_2 b_3}}{
c^{1/4}\left(a^2 e_0 - a e h_0 - \frac{c h_0}{3}\right)}  \ , \ \ \
\ k = 1,2,3\ , \ \ \ \ \ A_2^{44}= 0 \, .\eeq Since the null space
of $A_2$ is 1-dimensional,  supersymmetry is broken spontaneously to
${ \cal N } = 1$ by this vacuum.

\bigskip

\noindent
{\bf Type IIB isotropic AdS with only IIA-geometric fluxes}

Here we explore models without primed fluxes turned on, and without
the following fluxes
\begin{equation} h_0 = h_1 = q_2 = q_3 = q_6 = q_7 = p_i = 0 \, ,
\end{equation} which turn out to be T-dual to non-geometric fluxes in IIA.
The only fluxes we are left with are RR, NSNS and IIA-geometric
fluxes, leading to the superpotential
\begin{equation}
W = -f_3 + i h_3 S + 3 i q_1 T + 3 i f_2 U + 3 h_2 S U + 3 q_0 T U -
3 q_4 T U + 3q_5 TU + 3 f_1 U^2 - i f_0 U^3\, .
\label{spadsfixed}
\end{equation}
These models are not only the  simplest but also perhaps the
most interesting ones because all the ingredients  have a
well known interpretation in $D = 10$, and one could try to uplift
the solutions to fully consistent and under control supergravity
configurations.  The equations for fluxes  in  ${\cal N} = 4$ models
deformed by branes are
\begin{eqnarray}
N_{D3/O3}&=& N_{D3}- 16 = \frac{1}{2} \left[ f_0 h_3 - 3 f_1 h_2\right]\\
N_{D7} &=& \frac{1}{2}\left[f_0 q_1 - f_1 (q_0 - q_4 + q_5)\right] \\
0 &=& h_2 (q_4 - q_0) = h_2 (q_0 + q_5) = q_4^2  + q_0 q_5 = q_5^2 -
q_0 q_4
\end{eqnarray}
There are three independent solutions to these equations
\begin{enumerate}
\item\ \ \ \  $h_2 = \frac{N_{D3/O3} q_5}{3 N_{D7}} \ , \ \ \ f_0 = 0 \ , \ \ \ q_0 = q_4= - q_5 \ , \ \ \ f_1 = \frac{-2 N_{D7}}{q_5}$
\item\ \ \ \  $h_3 = \frac{3 f_1 h_2 + 2 N_{D3/O3}}{f_0} \ , \ \ \ q_1 = \frac{2 N_{D7} + f_1 q_5}{f_0} \ , \ \ \ q_0 = q_4= - q_5$
\item\ \ \ \  $h_3 = \frac{2 N_{D3/O3}}{f_0} \ , \ \ \ q_1 = \frac{2 N_{D7} + f_1 q_0}{f_0} \ , \ \ \ h_2 = q_4 = q_5 = 0$
\label{solucion}
\end{enumerate}
The deformed ${\cal N} = 8$ constraints, which for our specific
choice of fluxes read
\begin{equation} 0 = h_2 (q_5 - q_4)\, , \label{constN8}\end{equation} are only automatically satisfied by
solutions 3.

Here we would like to display an example of a fully stabilized AdS vacuum with
D3 and D7 branes, large volume and small coupling. It is not unique and
we only intend to show that string vacua of this kind exist. For
instance, solutions 2 lead in particular to the following vacuum
\begin{equation} s_x = - \frac{2 i \sqrt{f_1}\sqrt{f_3}}{3 h_2} \ , \ \ \ t_x =
-\frac{2 i \sqrt{f_1}\sqrt{f_3}}{q_5} \ , \ \ \ u_x = \frac{i
\sqrt{f_3}}{3 \sqrt{f_1}} \, ,\label{solAdS}\end{equation} with
vanishing axions $s_y = t_y = u_y = 0$. The brane charges in terms of the
fluxes read
\begin{eqnarray}
16  - N_{D3} &=& \left(\frac{3}{2} f_1 + \frac{5}{2}
\frac{f_2^2}{f_3}\right) h_2 \, ,\\
N_{D7} &=& -\left(\frac{1}{2} f_1 + \frac{5}{6}
\frac{f_2^2}{f_3}\right) q_5\, .
\end{eqnarray}
As we mentioned previously, due to the
 isotropic
symmetry, the tadpole cancelation conditions for D7 branes wrapping
the three possible four-cycles coincide. Here $N_{D7}$ indicates the
charge  of the D7 branes in each four-cycle. We also point out that this
flux configuration does not allow to include the S-dual NS7 and I7
branes. The volume and string coupling are given by
\begin{equation}
{\rm vol}_6 = \left(\frac{3 h_2}{q_5}\right)^{3/2} \ , \ \ \ \ \ g_S
= \frac{3 i h_2}{2 \sqrt{f_1}\sqrt{f_3}}
\end{equation}
Taking for example $f_3 = 1$, $q_5 = h_2 = 3$, $f_2 = 2$ and $f_1 =
-30$, one gets a configuration with $N_{D3} = 121$, $N_{D7} =
35$, ${\rm vol}_6 \sim 5$ and $g_S \sim 0.8$, and positive real parts of
all the moduli.

On the other hand, taking for example $f_1 = -15$, $f_2 = 3$, $f_3 =
q_5 = h_2 = 1$, one obtains a configuration with $N_{D3/O3} = N_{D7}= 0$,
${\rm vol}_6 \sim 5$ and $g_S \sim 0.4$. It  is interesting that
all brane
charges can vanish
because it signals the fact that this
is a vacuum of an exact truncation of gauged ${\cal N} = 4$
supergravity. Therefore, it could be conceived as an ${\cal
N}=4$ minimum spontaneously breaking ${\cal N} =4 \to 1$. In fact,
evaluating (\ref{solAdS}) in the shift matrix $A_2$, generically
yields
\begin{equation}
-i \frac{2}{3}e^{-K/2}A_{2}^{kk} =  \frac{8 f_2
\sqrt{f_3}}{9\sqrt{f_1}} -4 f_3  + \frac{4 \sqrt{f_1}f_3^{3/2}}{15
f_2}\ , \ \ \ \ \ \ k = 1,2,3
\end{equation}
and it is non vanishing when  the solution is plugged
in.

Imposing the ${\cal N} = 8$ constraint (\ref{constN8})
implies setting either $h_2$ or $q_5$ to zero, which is not allowed in
this solution and thus
 it does not correspond to an ${\cal N} = 8$
truncation. However,
if we take vanishing fluxes $f_1 = q_0 =q_4 = q_5 = h_2 = 0$
in the solutions  2 and 3,
equation (\ref{constN8}) is satisfied. In this case, for instance,
there is  a vacuum where all  moduli are fixed except for a combination
of axions, namely
\begin{equation}
s_x = - \frac{2\sqrt{5 f_2}}{3 \sqrt{3f_0}} \frac{f_2}{h_3} \ , \ \
\ \ t_x = - \frac{2\sqrt{5 f_2}}{3 \sqrt{3f_0}} \frac{f_2}{q_1} \ ,
\ \ \ \ u_x = \frac{\sqrt{5 f_2}}{\sqrt{3 f_0}}\ , \ \ \ \ u_y = 0 \
, \ \ \ \ h_3 s_y + 3 q_1 t_y + f_3 = 0 \, .\nn
\end{equation}
Then, assigning the values $f_2= -5$, $f_0= -1$, $q_1 = 2$ and $h_3
= 4$, one obtains a configuration with $N_{D3/O3} = -14$, $N_{D7} = -1$,
${\rm vol}_6 \sim 3$ and $g_S\sim 0.4$.
Given that in this vacuum
the net charge of 3 and 7-sources is non vanishing, the uplifting of this
configuration is expected to involve D3 and D7-branes as well as O3 and
O7-planes.
We then see in this
particular example a generic fact of these models: when ${\cal N} =
8$ constraints are imposed, many (would be) stabilized fields become
massless, but one can still find vacua satisfying all other
constraints.

\bigskip

\noindent {\bf Type IIB isotropic AdS with P-fluxes}

Models involving $P$-fluxes are interesting since they have an
interpretation in F-theory \cite{acr}. We have found fully
stabilized AdS vacua including
 these fluxes. For instance, taking
non vanishing fluxes $f_2,  q_1, h_2,h_3,p_0, p_1, p_4, p_5$, the
${\cal N}=4$ constraints are satisfied if $p_0=p_4=-p_5$. The
superpotential is given by
\begin{equation}
 W = 3 i (q_1 T + f_2 U) + S (i h_3 + 3 p_1 T + 3 h_2 U + 3 i p_0
TU) \, ,
\end{equation}
and it is easy to verify that all the moduli can be stabilized in
this model
 at values satisfying the physical constraints.

\subsubsection{Further constraints}
It is well known that, in presence of bulk fluxes, branes
could have non trivial FW anomalies \cite{Freed:1999vc}
and,
therefore, certain brane configurations are not allowed.
In a Type IIB setting, with magnetized branes,
 Chern-Simons couplings on the worldvolume of
the branes lead to the gauging of  certain RR axionic scalars under
D-brane $U(1)$ gauge symmetries.
In \cite{cfi} (see also
\cite{vz,
Villadoro:2007yq,oscar})
it was
suggested that FW constraints could be understood as the
requirements needed to ensure invariance of the effective
superpotential under shifts of the  $U(1)$ gauged axionic scalars. Also in
\cite{acr} (where only  magnetized 7-branes were considered), FW anomalies
were shown
to arise as JI of the algebra involving both, the closed
string  and the open string generators associated to $U(1)$ gauge
fields on the brane worldvolume.
 FW constraints must be taken into account when dealing with specific
models involving magnetized branes \cite{cu}.

Generically,
 the topological information of a set of $N_a$ D-branes is encoded in
six integers $(n_a^i,m_a^i)$:
$m_a^i$ is the number of times that the D-branes wrap the $i^{th}$ ${\rm T}^2$
and $n_a^i$ denotes the units
 of magnetic flux in that torus. This notation allows us
to describe different types of branes (D9, D7, etc.) \cite{Marchesano:2004yq}.
 Absence of anomalies imposes restrictions on these integers.

As an example, let us revisit the model (\ref{spadsfixed}) displayed
above. For a generic (non isotropic) superpotential, FW contraints
read  (see for instance \cite{cfi})
 \begin{eqnarray}
-h_0c_0^a+h_ic_i^a&=&0\, ,\\
a_ic_0^a+b_{ij}c_j^a&=&0\, ,
\end{eqnarray}
which in the isotropic case reduce to
\begin{eqnarray}
h_3c_0^a+q_1(c_1^a+c_2^a+c_3^a)&=&0\, ,\\
-h_2c_0^a-q_0c_1^a+q_4c_2^a-q_5c_3^a&=&0\, ,\\
-h_2c_0^a-q_5c_1^a-q_0c_2^a+q_4c_3^a&=&0\, ,\\
-h_2c_0^a+q_4c_1^a-q_5c_2^a-q_0c_3^a&=&0\, ,
\end{eqnarray}
with
$c_0^a= m_a^1 m_a^2 m_a^2,\, c_1= m_a^1 n_a^2 n_a^2,\, c_2= n_a^1 m_a^2
n_a^2,\, c_3= n_a^1 n_a^2 m_a^2$.

Using the conditions  $q_0 = q_4= - q_5$  for solutions 1 and 2 above, we
easily see that
$q_0\ne 0$ implies
\begin{eqnarray}
c_1^a&=&c_2^a=c_3^a =c^a\, ,\\
\frac{c_0^a}{c^a}&=&\frac{q_0}{h_2}=-\frac{3q_1}{h_3}\, ,
\end{eqnarray}
for every stack of $N^a$ branes.
Interestingly enough, when looking at the ``intersection number'' of two sets
$a$ and  $b$ of D-branes, namely
\beq
I_{ab} = \prod_{i=1}^{3} \left(n_a^i m_b^i - m_a^i n_b^i \right)\, ,
\label{intersection}
\eeq
we find that these strong constraints lead to $I_{ab}=0$,
the spectrum being non
chiral. A similar situation arises for solution 3. Here if $q_0\ne 0$ then all
coefficients $c^a$ must vanish.
Thus, we conclude that a chiral spectrum requires $q_0=0$, and in this case
we have seen that some
linear combination of axion fields remains unfixed.

Actually, this at first sight unpleasant result is useful. As noted
in \cite{cfi} (in the Type IIA picture), these axions could be
helpful to eliminate anomalous $U(1)$  when looking for a chiral
spectrum.


\subsubsection{Minkowski}

{\bf No fully stabilized isotropic Minkowski vacuum with only $F, H, Q$ and
$P$ fluxes}

Here we show that generically, fully stabilized Minkowski vacua
cannot be achieved in the absence of primed fluxes\footnote{In orientifolds with SU(3)$\times$SU(3) structure,
it was shown in \cite{micu} that it is impossible to stabilize all
the moduli in a supersymmetric Minkowski vacuum without non-geometric fluxes. }. In \cite{acfi}, simple models of Minkowski
vacua were found with $F, H, Q$ and $P$ fluxes, which were only
constrained by JI in the presence of branes. Other restrictions such
as the antisymmetric constraints were not considered, and it can be
checked that those conditions can only be satisfied for non vanishing
$K^{ABCD}$. Therefore, one expects the $D = 10$ uplifting of those
vacua to involve exotic objects \cite{Villadorosource}. In the case
that we are considering, we now show that it is not possible to
stabilize all moduli with non-vanishing real components in a
Minkowski vacuum  when only $F, H, Q$ and $P$ fluxes are present.
Clearly, since a Minkowski vacuum was found in \cite{acfi}, the
proof must involve both the constraints and the structure of the
superpotential. The only three facts needed to
prove our statement are:

\begin{enumerate}
\item One can explicitly see in figure \ref{Figura} that the antisymmetric
constraints  $q_2=q_3=q_4=q_5= p_2 = p_3 = p_4 = p_5 = 0$ must
generically be imposed to find SUSY Mkw vacua.
\item The superpotential has the structure
\begin{equation}
W = F(U) + H(U) S + Q(U) T + P(U) S T
\end{equation}
and the Minkowski SUSY equations are
\begin{eqnarray}
H(U) + P(U) T &=& 0 \label{uno}
\\
Q(U) + P(U) S &=& 0 \label{dos} \\
F(U) + Q(U) T = F(U)+ H(U) S &=&0 \label{tres} \\
F'(U) + H'(U) S + Q'(U) T + P'(U) S T &=&0 \label{cuatro}
\end{eqnarray}
 Since $T,S\neq 0$, one can verify from
(\ref{uno})-(\ref{tres}) that in the vacuum either $F, H, Q$ and $P$
all vanish or are  non zero. In the former case, only one
equation (\ref{cuatro}) is left to stabilize $T$ and $S$, so there
will necessarily be at least one modulus. Therefore, all terms should be non
vanishing.
\item $Q(U)$ and $P(U)$ are given by
\begin{equation}
Q(U) = 3i (q_1 -i  q_0 U - q_6 U^2 + i q_7 U^3) \ ,  \ \ P(U)= 3(
p_1 -
 i p_0 U -   p_6 U^2 + i p_7 U^3).
\end{equation}
When only two terms of each -differing by a power of $U$-  are
present, and the parameters of $Q$ are proportional to those of $P$,
it is impossible to stabilize the real part of some modulus away
from the origin. For example in the case  $q_0 = q_1 = p_0 = p_1 =
0$ one can verify that when $p_6 / q_6 = p_7 / q_7$,
\begin{equation}
\partial _T W = Q(U) +  S P(U) = \frac{3}{q_7} U^2 (-q_6 + i q_7 U) (p_7  S + i
 q_7) = 0
\end{equation}
and then, the real part of $S$ or $U$ is forced to vanish. This is
trivially generalized to all the cases.
\end{enumerate}
Combining these three facts it can readily be shown that it is
impossible to fully stabilize vacua in Minkowski with all real moduli
non-vanishing. In fact, when primed fluxes are set to zero,
together with the fluxes of point 1, it is easy to verify that the
solutions to the constraints impose that either $F, H, Q$ or $P$
vanish, or $Q$ and $P$ fall in the case mentioned in point
3.\footnote{There is a unique model that avoids this exclusion
argument, but it does not stabilize the moduli properly.}

This of course does not mean that there are no Minkowski vacua, and
in fact there are plenty degenerate vacua of this type. To look for
partially stabilized susy Minkowski minima, it is convenient to
consider only non-primed electric gaugings. This implies, on the one
hand, that the constraints (\ref{Singlet}) and (\ref{FHCov}) are
automatically satisfied, and on the other, that the vacuum will
necessarily be Minkowski and have a continuous degeneracy in the
axion-dilaton direction. We have pointed out that due to the
structure of the superpotential, in general the fields are
stabilized at purely imaginary values. To overcome this difficulty,
we set to zero all the terms involving an even number of fields.
This
 ensures that all the parameters in the superpotential are real, and
allows for real VEV  of the K\"ahler and complex structure moduli.
After imposing the constraints we are left with the following
superpotential \beq W=3 f_2 U - f_0 U^3 + T (3q_1 - 3 q_3 U^2)\, .
\eeq

This model has D7-branes  and vanishing $N_{D3/O3}$
\begin{equation}
N_{D7} = \frac{1}{2}\left[f_0q_1+f_2q_3\right] \ , \ \ \ \ N_{D3/O3} = 0\, .
\end{equation}
In particular, for $f_0=3f_2q_3/q_1$, it admits  minima with the
following moduli \beq T=-\frac{f_2}{\sqrt{q_1}\sqrt {q_3}}\, , \ \ \
U=\frac {\sqrt{q_1}}{\sqrt{q_3}}\, , \ \ \ \forall S\, . \eeq As we
anticipated, since $W$ does not depend on $S$, its Kahler
derivatives with respect to $S$ are proportional to the
superpotential itself, so every SUSY vacua is necessarily Minkowski.
Let us finally point out that this solution can lie in the large
volume and small coupling regimes by suitable choices of the
parameters, and consistency of the solution requires non vanishing
$N_{D7}$.

\section{Summary and outlook.} \label{conc}

Compactifications of string theory to $D=4$ can preserve
 different amounts of supersymmetries. The paradigm of
low scale physics has been to construct
 ${\cal N} = 1$  effective theories in
which supersymmetry is  spontaneously or dynamically
 broken at low energies. This can be realized in
 compactifications on Calabi-Yau manifolds, or
alternatively on spaces with reduced holonomy in which the large
supersymmertic structure
is explicitly  broken by projections.
 In this  case, the effective theory is
a truncation of the parent theory, and one expects that it  inherits many
of the original characteristic features.

In this work we have focused on Type IIB toroidal compactifications
in which supersymmetry is broken from ${\cal N} = 8$ to ${\cal N}=4$
by an orientifold projection incorporating O3 planes.
In addition, supersymmetry is further broken to ${\cal N} = 1$ by a
$\mathbb{Z}_2 \times \mathbb{Z}_2$ orbifold projection, where
the structure group is enhanced to $SU(3)$ at the fixed points. The
untwisted sector of the resulting theory is given by direct
truncation of the parent ${\cal N}=4$ theory, so it is expected to match
some ${\cal N} = 4$ gauged
supergravity truncation.
We have clarified the connection  between
these string orbifolds and gauged supergravity truncations.

We  established the full dictionary
between the scalar sector of gauged supergravity and
the moduli
space of the string orbifold (\ref{Mi}) and, by extending the results of
\cite{acr}, we determined  the complete connection between
fluxes and gaugings (\ref{dic}). Then, we  analyzed the SUSY
variations of the fermions in gauged supergravity, which are
parameterized by the so-called shift matrices (\ref{shiftmat}).
These matrices fix the structure of the scalar potential
(\ref{scag}) and the amount of supersymmetry preserved by the vacuum
(\ref{spontSUSYbreak}). We  applied the orbifold projection on
the shift matrices and recovered the effective Type IIB
superpotential (\ref{superpotgeneral}) and the scalar potential
obtained from it. This complements the results of \cite{acr}, where
 the complete set of constraints on fluxes in the string side
were shown to exactly match the gauged supergravity consistency
conditions for gaugings (\ref{ji}).

We have extended the analysis so as to include branes in this setup.
A brief computation  shows that when these sources are present, the
structure of the effective scalar potential remains unaltered in the
untwisted sector. However, branes source the tadpole cancelation
conditions of string theory, and in many cases break the exact
matching with the quadratic constraints of gauged supergravity. Such
is the case of D7 branes, which project out a different combination
of gravitini than the orientifold planes and break ${\cal N} = 4$
 explicitly. Therefore, even if the effective theory
{\it looks like} a gauged supergravity truncation in the
presence of branes,
strictly speaking it is not. Rather, it is a deformation
that corresponds to an exact truncation when no SUSY-breaking
sources are present.

We have also obtained additional constraints on fluxes by comparing
the scalar potentials of gauged ${\cal N} = 4$ and ${\cal N}=8$
supergravities (\ref{Singlet})-(\ref{FHCov}). Such constraints
include the tadpole cancelation condition stating that no D3/O3
sources or any of their duals should be present.
From a $D = 4$
perspective, such constraints seem to correspond to a cancelation of the
scalar masses required by ${\cal N} = 8$ consistency.

Having collected the full set of constraints that one expects in
these string truncations, we addressed the problem of moduli fixing.
The rich structure of the superpotential (\ref{superpotgeneral}) is
highly constrained by the consistency requirements. We have seen
that in most models the constraints force a proportionality
between electric and magnetic gaugings, and showed that it is not
possible to stabilize all moduli in such case. With this argument, we
were able to exclude large regions of the parameter space and focus on
models which have potentiality for full moduli fixing. We displayed some
examples of AdS and Minkowski SUSY vacua, analyzed their
${\cal N} = 4, 8$ origin, and also discussed the possibility that they
spontaneously break ${\cal N} = 4 \to 1$. Some of the examples
have all moduli stabilized in a vacuum with (or without) branes in large
compactification volume and small coupling regimes.
In a subset of  AdS vacua, we have explicitly shown that, as observed in
\cite{cfi},
a chiral spectrum is connected with the presence of unfixed axions.

\medskip
{\bf Note added}: Soon after this paper appeared in the arXiv, we
 received the preprint \cite{guar}, having some overlap with parts of
our work. See also the more recent paper \cite{Dibitetto:2011eu}.

\bigskip

{\bf \large Acknowledgments} { We thank E. Andr\'es, G. Dall'Agata,
 G. A. Guarino and H. Samtleben for useful
discussions, comments and correspondence, and especially P. C\'amara
and M. Gra\~na, for also suggesting important improvements on the
manuscript. We are grateful to D. Mitnik for providing access to his
computer network. This work was partially supported by MINCYT
(Ministerio de Ciencia, Tecnolog\'\i a e Innovaci\'on Productiva of
Argentina) and ECOS-Sud France binational collaboration project
A08E06. D.M. benefited from the CNOUS/CONICET Bernardo Houssay
Fellowship and ANR grant 08–JCJC–0001–0.}

\appendix

\section{Constraints and superpotential.} \label{app}

Here we would like to compile the explicit form of the constraints
and superpotential, and link the different notations used
for the fluxes and the parameters.

The gauge algebra has the form
\begin{center}
\begin{tabular}{ccc}
$\ \ \;\; \left[X^a, X^b\right] \ \ =\ \ -\ti F^{abp}Z_p + Q^{ab}_p
X^p$ & $\ \ \ $ & $ \ \ \ \; \left[\bar X^a, \bar X^b\right] \ \ =\
\ -\ti H^{abp}\bar Z_p + P^{ab}_p
\bar X^p$ \\
$\left[X^a, Z_b\right] \ \   =\ \   Q'^a_{bp} X^p - Q^{ap}_b Z_p$ &
$\ \ \ $& $\left[\bar X^a, \bar Z_b\right] \ \ =\ \ P_{bp}'^a \bar
X^p - P^{ap}_b
\bar Z_p$\\
$\ \ \left[Z_a, Z_b\right] \ \  =\ \  - F_{abp}' X^p +  Q_{ab}'^p
Z_p$ & $\ \ \ $& $\ \ \; \left[\bar Z_a, \bar Z_b\right] \ \ =\ \  -
H_{abp}' \bar X^p + P_{ab}'^p \bar Z_p$
\end{tabular}
\end{center}
\beqa &&\left[X^a, \bar X^b\right]\ \  =\ \  Q^{ab}_p \bar X^p -
\tilde F^{abp} \bar Z_p
\ \  =\ \ P_p^{ab}  X^p - \tilde H^{abp}  Z_p\nn\\
\left[\bar X^a,  Z_b\right]&=&\left[X^a, \bar Z_b\right] \ \  =\ \
Q'^{a}_{bp} \bar X^p - Q^{ap}_b \bar Z_p
\ \  =\ \ P'^{a}_{bp} X^p -P^{ap}_b  Z_p \nn\\
&&\left[Z_a, \bar Z_b\right] \ \  =\ \  -F'_{abp} \bar X^p +
Q'^p_{ab} \bar Z_p \ \  =\ \  P'^{p}_{ab}Z_p-H'_{abp}X^p\,
\label{Algebra} \eeqa
The last equalities are enforced to ensure antisymmetry of the
commutators, and imply the following ``antisymmetry constraints''
for fluxes (\ref{ji})
 \beqa Q_{ab}'^p P_{cp}'^l - P_{ab}'^p Q_{cp}'^l
+F'_{pab}P^{lp}_c -H'_{pab}
Q^{lp}_c &=& 0\nn\\
P_{ab}'^p F_{clp}' -  Q_{ab}'^p H'_{clp}
-F'_{pab}P_{cl}'^{p}+H'_{pab}
Q'^{p}_{cl}&=& 0\nn\\
 Q_{ab}'^p P_p^{cl} - P_{ab}'^p Q_p^{cl}+F'_{pab}\tilde H^{clp}-H'_{pab}
\tilde F^{clp} &=& 0\nn\\
Q^{pb}_a P_{cp}'^l - P_a^{pb} Q_{cp}'^l + Q'^b_{pa}
P^{lp}_c-P'^b_{pa}
Q^{lp}_c &=&0\nn\\
Q^{pb}_a P^{cl}_p - P^{pb}_a  Q_p^{cl} + Q '^b_{pa}\tilde
H^{clp}-P'^b_{pa}
\tilde F^{clp}&=& 0 \nn\\
\ti F^{pab} P^{cl}_p - \ti H^{pab} Q_p^{cl}+Q^{ab}_p\tilde
H^{clp}-P^{ab}_p \tilde F^{clp} &=& 0\label{AntiSymmetry} \eeqa
Using these equations, the Jacobi identities for the full set of
fluxes can be written as
\begin{center}
\begin{tabular}{ccc}
$Q_{[a b}'^{p}  Q_{c]p}'^{l} + F'_{p [a b} Q^{l p}_{c]}\ \ =\ \ 0$ &
$\ \
\ $ & $Q_{[a b}'^{p} F_{c]lp}' + F'_{p [a b}  Q_{c] l}'^{p}\ \ =\ \ 0$ \\
$Q_{[a b}'^{p} P_{c]p}'^{l} + F'_{p [a b} P^{l p}_{c]}\ \ =\ \ 0$ &
$\ \
\ $& $Q_{[a b}'^{p} H_{c]lp}' + F'_{p [a b} P_{c] l}'^{p}\ \ =\ \ 0$\\
$P_{[a b}'^{p} P_{c]p}'^{l} + H'_{p [a b} P^{l p}_{c]}\ \ =\ \ 0$ &
$\ \ \ $& $P_{[a b}'^{p} H_{c]lp}' + H'_{p [a b} P_{c] l}'^{p}\ \ =\
\ 0$
\end{tabular}
\end{center}
\beqa F_{pab}' \ti F ^{clp} +  Q_{ab}'^p Q^{cl}_p - Q^{pc}_{[a}
Q_{b]p}'^l -
Q_{p[a}'^c Q_{b]}^{lp}&=& 0 \nn\\
F_{pab}' \ti H ^{clp} +  Q_{ab}'^p P^{cl}_p - Q^{pc}_{[a} P_{b]p}'^l
-
Q_{p[a}'^c P_{b]}^{lp} &=& 0 \label{Jacobis}\\
H_{pab}' \ti H ^{clp} + P_{ab}'^p P^{cl}_p - P^{pc}_{[a} P_{b]p}'^l
- P_{p[a}'^c P_{b]}^{lp} &=& 0 \nn \eeqa
\begin{center}
\begin{tabular}{ccc}
$Q_{p}^{[bc} Q^{l] p}_a -  Q_{pa}'^{[b} \ti F^{c l] p} \ \ =\ \  0$
& $\ \
\ $ & $\ti F^{p[ab} Q^{c] l}_p + Q^{[ab}_p \ti F^{c] l p} \ \ =\ \  0$ \\
$P^{[bc}_p Q^{l]p}_a -  Q_{pa}'^{[b}\ti H^{cl]p} \ \ =\ \  0$ & $\ \
\ $&
$\ti F^{p[ab} P^{c] l}_p + Q^{[ab}_p \ti H^{c] l p} \ \ =\ \  0$\\
$P_{p}^{[bc} P^{l] p}_a - P_{pa}'^{[b} \ti H^{c l] p} \ \ =\ \  0$ &
$\ \ \ $& $\ti H^{p[ab} P^{c] l}_p + P^{[ab}_p \ti H^{c] l p} \ \ =\
\  0$
\end{tabular}
\end{center}
The superpotential explicitly reads
\beqa
W & = &
e_0-i\sum_{i=1}^3 h_iT_i+\frac{1}{2}\sum_{l\aneq m\aneq
n} h'_lT_mT_n+i e'_0 T_1T_2T_3 \nn \\
& + & \bigg(ih_0-\sum_{i=1}^3 f_iT_i-\frac{i}{2}\sum_{l\aneq
m\aneq
n} f'_lT_mT_n - h'_0 T_1T_2T_3 \bigg)S \nonumber \\
& + & \sum_{i=1}^3 \left[\bigg(-a_i+i\sum_{j=1}^3g_{ij}T_j
-\frac{1}{2}\sum_{l\aneq m\aneq n} g'_{il}T_mT_n + ia'_i T_1T_2T_3 \bigg)S
\right. \nonumber \\
& + & \left.  ie_i-\sum_{j=1}^3 b_{ij}T_j-\frac{i}{2}\sum_{l\aneq
m\aneq n}b'_{il}T_mT_n
-e'_iT_1T_2T_3 \right] U_i \nonumber\\
& + &  \frac{1}{2}\sum_{r\aneq s\aneq
t}\left[\bigg(i\bar{a}_r+\sum_{j=1}^3\bar{g}_{rj}T_j+\frac{i}{2}
\sum_{l\aneq m\aneq n} \bar{g}'_{rl}T_mT_n - \bar{a}'_rT_1T_2T_3 \bigg)S \right.
\nonumber\\
& - &  \left. q_r+i\sum_{j=1}^3\bar{b}_{rj}T_j
-\frac{1}{2}\sum_{l\aneq m\aneq n}\bar{b}'_{rl}T_mT_n + iq'_r
T_1T_2T_3\right]U_sU_t \nonumber \\
& + &
\left[-\bigg(\bar{h}_0+i\sum_{j=1}^3\bar{f}_{j}T_j-\frac{1}{2}\sum_{l\aneq
m\aneq n}\bar{f}'_lT_mT_n + i\bar{h}'_0T_1T_2T_3\bigg)S \right. \nonumber\\
& + &  \left. im+\sum_{j=1}^3\bar{h}_{j}T_j+\frac{i}{2}\sum_{l\aneq
m\aneq n}\bar{h}'_lT_mT_n- m'T_1T_2T_3\right]U_1U_2U_3\, .
\label{sl}
\eeqa
The connection between these parameters
and the IIB fluxes  can be read from Table 1. We also refer to the
original literature \cite{acfi, acr} for additional details
on the notation, in particular for links with IIA and Type I fluxes.

Finally, we display  the ${\cal N} = 8$ constraints (for simplicity
when only $F, H, Q$ and $P$ fluxes are turned on)
\begin{equation}
\tilde F^{pqr}H_{pqr}
 = 0\ , \ \ \ \ H_{pqa}Q^{pq}_b - F_{pqa} P^{pq}_b = 0 \ , \ \ \ \ \hat \sigma_{pq} P^{p[a}_{[c} Q^{b]q}_{d]} = 0 \ , \ \ \ \ \hat \sigma = i \sigma^2 \otimes 1_3\end{equation}

\section{Useful formulas. }

Vector indices of $SU(4)$ are raised and lowered
 with complex conjugation, as $v_i=(v^i)^{*}$. On the other
hand,  $SO(6)$ vectors can be conveniently described through
antisymmetric tensors $\nu^{i j}$ subject to the pseudo-reality
constraint
\begin{equation}
 \nu_{ij}=(\nu^{ij})^{*}=\frac{1}{2}\epsilon_{ijkl}\nu^{kl}\, ,
\end{equation}
with scalar product
\begin{equation}
 \nu_m\nu_m=\frac{1}{2}\epsilon_{ijkl}\nu^{ij}\nu^{kl}\, .
\end{equation}
Being $\{e^1,\dots,e^6\}$ the canonical basis of $SO(6)$, then the
corresponding basis of $SU(4)$ is given by
\begin{eqnarray}
&& e^1 \leftrightarrow (\gamma^1)^{ij} = \frac{i}{2}  \sigma^2
\otimes \sigma^1 \ , \ \ \ \ e^2 \leftrightarrow (\gamma^2)^{ij} =
-\frac{i}{2}  \sigma^2 \otimes \sigma^3 \ , \ \ \ \ e^3
\leftrightarrow (\gamma^3)^{ij} = \frac{i}{2}  1 \otimes
\sigma^2 \, ,\nn\\
&& e^4 \leftrightarrow (\gamma^4)^{ij} = -\frac{1}{2}  \sigma^1
\otimes \sigma^2 \ , \ \ \ \ e^5 \leftrightarrow (\gamma^5)^{ij} =
-\frac{1}{2}  \sigma^1 \otimes 1 \ , \ \ \ \ e^6 \leftrightarrow
(\gamma^6)^{ij} = \frac{1}{2}  \sigma^3 \otimes \sigma^2 \, .\nn
\end{eqnarray}
And we can rewrite the coset representative as
\begin{equation}
 \nu^M=\nu_m^{\quad M}e^m \leftrightarrow \nu_m^{\quad M}(\gamma^m)^{ij}\, .
\end{equation}

Also the basis matrices have the following  useful properties
\begin{eqnarray}
 {\rm Tr} [\gamma_{i_1} \gamma^*_{i_2}\gamma_{i_3}
\gamma^*_{i_4}] &=& \frac{1}{4} (\delta_{i_1i_2}\delta_{i_3i_4} -
\delta_{i_1 i_3}\delta_{i_2i_4}+\delta_{i_1i_4}\delta_{i_2i_3})\, ,\\
{\rm Tr} [\gamma_{i_1}^* \gamma_{i_2}\gamma_{i_3}^*
\gamma_{i_4}\gamma_{i_5}^* \gamma_{i_6}] &=& - \frac{i}{16}
\epsilon_{i_1i_2i_3i_4i_5i_6}  - \frac{1}{16} [
\delta_{i_1i_2}\delta_{i_3i_4}\delta_{i_5i_6} -
\delta_{i_1i_2}\delta_{i_3i_5}\delta_{i_4i_6} +
\delta_{i_1i_2}\delta_{i_3i_6}\delta_{i_4i_5} - \nn \\
&&\ \ \ \ \ \ \ \ \ \ \ \ \ \ \ \ \ \ \ \ \ \ \ \ \ \ \
\delta_{i_1i_5}\delta_{i_3i_4}\delta_{i_2i_6} +
\delta_{i_1i_6}\delta_{i_3i_4}\delta_{i_2i_5} -
\delta_{i_1i_3}\delta_{i_2i_4}\delta_{i_5i_6} + \nn \\
&&\ \ \ \ \ \ \ \ \ \ \ \ \ \ \ \ \ \ \ \ \ \ \ \ \ \ \
\delta_{i_1i_4}\delta_{i_2i_3}\delta_{i_5i_6} +
\delta_{i_1i_3}\delta_{i_2i_5}\delta_{i_4i_6} -
\delta_{i_1i_3}\delta_{i_2i_6}\delta_{i_4i_5} - \nn \\
&&\ \ \ \ \ \ \ \ \ \ \ \ \ \ \ \ \ \ \ \ \ \ \ \ \ \ \
\delta_{i_1i_5}\delta_{i_2i_3}\delta_{i_4i_6} +
\delta_{i_1i_6}\delta_{i_2i_3}\delta_{i_4i_5} -
\delta_{i_1i_4}\delta_{i_2i_5}\delta_{i_3i_6} + \nn \\
&&\ \ \ \ \ \ \ \ \ \ \ \ \ \ \ \ \ \ \ \ \ \ \ \ \ \ \
\delta_{i_1i_4}\delta_{i_2i_6}\delta_{i_3i_5} +
\delta_{i_1i_5}\delta_{i_2i_4}\delta_{i_3i_6} -
\delta_{i_1i_6}\delta_{i_2i_4}\delta_{i_3i_5} ]\, ,\nonumber
\end{eqnarray}
that allow to link the expressions (\ref{scag}) and
(\ref{massmatrix}) of the scalar potentials.

~
\newpage
~

\begin{table}[!ht] \footnotesize
\renewcommand{\arraystretch}{1.25}
\begin{center}
\caption{Dictionary between the different notations adopted for the
flux parameters.}
\begin{tabular}{ccc|ccc|ccc|ccc}
$F$ & param & $\mathbb{Z}_3$ & $H$ & param &  $\mathbb{Z}_3$&
$F'$ & param & $\mathbb{Z}_3$
&$H'$ & param & $\mathbb{Z}_3$\\
\hline $F_{123}=-\tilde{F}^{456}$ & $-m$ & $f_0$ & $H_{123}$ & $\ \
\bar h_0$ & $h_0$
& $F'_{123}$ & $\ \ m'$ &$f'_0$& $H_{123}'$ & $- \bar h_0 '$ & $h'_0$\\
$F_{423}=\tilde{F}^{156}$ & $-q_1$ & $f_1$ &$H_{423}$ & $-\bar a_1$
&$h_1$&
$F'_{423}$ & $- q_1'$ &$f'_1$& $H_{423}'$ & $\ \ \bar a_1'$ & $h'_1$\\
$F_{153}=\tilde{F}^{426}$ & $-q_2$ &$f_1 $& $H_{153}$ & $-\bar a_2$ & $h_1 $&
$F'_{153}$ & $- q_2'$ &$f'_1 $   &$H_{153}'$ & $\ \ \bar a_2'$ &$h'_1 $ \\
$F_{126}=\tilde{F}^{453}$ & $-q_3$&$f_1 $ & $H_{126}$ & $-\bar a_3$ &$h_1 $ &
$F'_{126}$ & $- q_3'$ & $f'_1 $  & $H_{126}'$ & $\ \ \bar a_3'$& $h'_1 $\\
$F_{156}=-\tilde{F}^{423}$ & $\ \ e_1$ & $f_2 $& $H_{156}$ & $-a_1$&$h_2$
&
$F'_{156}$ & $- e_1'$ &$f'_2$ &$H_{156}'$ & $-a_1'$&$h'_2$\\
$F_{426}=-\tilde{F}^{153}$ & $\ \ e_2$ &$f_2 $& $H_{426}$ & $-a_2$ &$h_2 $&
$F'_{426}$ & $- e_2'$ & $f'_2 $ & $H_{426}'$ & $-a_2'$&$h'_2 $\\
$F_{453}=-\tilde{F}^{126}$ & $\ \ e_3$ &$f_2 $& $H_{453}$ & $-a_3$ &$h_2 $&
$F'_{453}$ & $- e_3'$ &$f'_2 $& $H_{453}'$ & $-a_3'$&$h'_2 $\\
$F_{456}=\tilde{F}^{123}$ & $-e_0$ &$f_3$& $H_{456}$ & $\ \ h_0$ &$h_3$&
$F'_{456}$ & $\ \ e_0'$ &$f'_3$& $H_{456}'$ & $- h_0'$&$h'_3$
\end{tabular}

\begin{tabular}{ccc}
$Q$ & param &$\mathbb{Z}_3$ \\
\hline $\bmat{ccc} \! \! \! Q^{23}_4 & Q^{31}_5 & Q^{12}_6 \! \!
\!\emat$ & $ -\bmat{ccc} \! \! \!  h_1 & h_2 & h_3  \! \! \!\emat$ &
$q_1$
\\[0.35cm]

$\bmat{ccc} \! \! \!
-Q^{23}_1 & \, Q^{34}_5 & \, Q^{42}_6 \\
\, Q^{53}_4 & \! \! \! -Q^{31}_2 & \, Q^{15}_6 \\
\, Q^{26}_4 & \, Q^{61}_5 & \! \! \! -Q^{12}_3 \emat $ & $\bmat{ccc} b_{11} &
b_{12} & b_{13} \\
b_{21} & b_{22} & b_{23} \\
b_{31} & b_{32} & b_{33} \emat$ & $\begin{matrix} b_{ii} = -q_0
\\ b_{12}=b_{23}=b_{31} =  q_4 \\ b_{13}= b_{21}=b_{32}=-q_5\end{matrix}$
\\[0.45cm]

$\bmat{ccc} \! \! \! Q^{56}_1 & Q^{64}_2 & Q^{45}_3  \! \! \! \emat$
& $ \! \! -\bmat{ccc} \! \! \!  \bar h_1 & \bar h_2 & \bar h_3  \!
\! \! \emat$ &$ q_7$\\[0.35cm]

$\bmat{ccc} \! \! \!
-Q^{56}_4 & \, Q^{61}_2 & \, Q^{15}_3 \\
\, Q^{26}_1 & \! \! \! -Q^{64}_5 & \, Q^{42}_3 \\
\, Q^{53}_1 & \, Q^{34}_2 & \! \! \! -Q^{45}_6 \emat$ & $\bmat{ccc}
\bar b_{11} & \bar b_{12} & \bar b_{13} \\
\bar b_{21} & \bar b_{22} & \bar b_{23} \\
\bar b_{31} & \bar b_{32} & \bar b_{33} \emat$&$\begin{matrix} \bar
b_{ii} = -q_6
\\ \bar b_{12} = \bar b_{23}=\bar b_{31}=-q_3 \\
\bar b_{13}= \bar b_{21} = \bar b_{32}= q_2\end{matrix}$
\end{tabular}

\begin{tabular}{ccc}
$ Q '$ & param &$\mathbb{Z}_3$\\
\hline $\bmat{ccc} \! \! \!  Q_{56}'^1 &  Q_{64}'^2 &  Q_{45}'^3 \!
\! \!\emat$ & $ -\bmat{ccc} \! \! \!  h_1' & h_2' & h_3'  \! \!
\!\emat$ & $q'_7$ \\[0.35cm]

$\bmat{ccc} \! \! \!
- Q_{65}'^4 & \,  Q_{16}'^2& \,  Q_{51}'^3 \\
\,  Q_{62}'^1 & \! \! \! - Q_{46}'^5 & \,  Q_{24}'^3 \\
\,  Q_{35}'^1 & \,  Q_{43}'^2 & \! \! \! - Q_{54}'^6 \emat $ & $\bmat{ccc}
b_{11}' & b_{12}' & b_{13}' \\
b_{21}' & b_{22}' & b_{23}' \\
b_{31}' & b_{32}' & b_{33}' \emat$ & $\begin{matrix} b'_{ii} = \ \
q'_6
\\  b'_{12} = b'_{23}=b'_{31}= q'_3 \\ b'_{13}=b'_{21}=
b'_{32}= - q'_2\end{matrix}$\\[0.45cm]

$\bmat{ccc} \! \! \!  Q_{32}'^4 &  Q_{13}'^5 &  Q_{21}'^6  \! \! \!
\emat$ & $ \! \! -\bmat{ccc} \! \! \!  \bar h_1' & \bar h_2' & \bar
h_3'  \! \! \! \emat$ & $-q'_1$ \\[0.35cm]

$\bmat{ccc} \! \! \!
- Q_{23}'^1 & \,  Q_{34}'^5 & \,  Q_{42}'^6 \\
\,  Q_{53}'^4 & \! \! \! - Q_{31}'^2 & \,  Q_{15}'^6 \\
\,  Q_{26}'^4 & \,  Q_{61}'^5 & \! \! \! - Q_{12}'^3 \emat$ & $\bmat{ccc}
\bar b_{11}' & \bar b_{12}' & \bar b_{13}' \\
\bar b_{21}' & \bar b_{22}' & \bar b_{23}' \\
\bar b_{31}' & \bar b_{32}' & \bar b_{33}' \emat$ &$\begin{matrix}
\bar b'_{ii} = -q'_0
\\ \bar b'_{12} = \bar b'_{23}=\bar b'_{31}= q'_4 \\ \bar b'_{13}=
\bar b'_{21}=\bar b'_{32}=-q'_5\end{matrix}$
\end{tabular}
\end{center}
\end{table}

\begin{table}[!ht] \footnotesize
\renewcommand{\arraystretch}{1.25}
\begin{center}
\begin{tabular}{ccc}
 $P$ & param &${\mathbb Z}_3$\\
\hline $\bmat{ccc} \! \! \! P^{23}_4 & P^{31}_5 & P^{12}_6 \! \!
\!\emat$ & $- \bmat{ccc} \! \! \!  f_1 & f_2 & f_3  \! \! \!\emat$
& $p_1$\\[0.35cm]

$\bmat{ccc} \! \! \!
-P^{23}_1 & \, P^{34}_5 & \, P^{42}_6 \\
\, P^{53}_4 & \! \! \! -P^{31}_2 & \, P^{15}_6 \\
\, P^{26}_4 & \, P^{61}_5 & \! \! \! -P^{12}_3 \emat $ & $\bmat{ccc} g_{11} &
g_{12} & g_{13} \\
g_{21} & g_{22} & g_{23} \\
g_{31} & g_{32} & g_{33} \emat$& $\begin{matrix} g_{ii}=-p_0\\
g_{12}=g_{23}=g_{31}=p_4\\g_{13}=g_{21}=g_{32}=-p_5\end{matrix}$\\
[0.45cm]
 $\bmat{ccc} \! \! \! P^{56}_1 & P^{64}_2 &
P^{45}_3 \! \! \! \emat$ & $ \! \! -\bmat{ccc} \! \! \!  \bar f_1 &
\bar f_2 & \bar f_3  \! \! \!
\emat$& $p_7$  \\[0.35cm]

 $\bmat{ccc} \!
\! \!
-P^{56}_4 & \, P^{61}_2 & \, P^{15}_3 \\
\, P^{26}_1 & \! \! \! -P^{64}_5 & \, P^{42}_3 \\
\, P^{53}_1 & \, P^{34}_2 & \! \! \! -P^{45}_6 \emat$ & $\bmat{ccc}
\bar g_{11} & \bar g_{12} & \bar g_{13} \\
\bar g_{21} & \bar g_{22} & \bar g_{23} \\
\bar g_{31} & \bar g_{32} & \bar g_{33} \emat$ & $\begin{matrix}
\bar g_{ii}=-p_6\\ \bar g_{12}=\bar g_{23}=\bar g_{31}=-p_3
\\ \bar g_{13}=\bar g_{21}
=\bar g_{32}=p_2
\end{matrix}$
\end{tabular}

\begin{tabular}{ccc}
 $P'$ & param & ${\mathbb Z}_3$ \\
\hline $\bmat{ccc} \! \! \! P_{56}'^1 & P_{64}'^2 &
P_{45}'^3 \! \!
\!\emat$ & $- \bmat{ccc} \! \! \!  f_1' & f_2' & f_3'  \! \! \!\emat$
& $p'_7$\\[0.35cm]

 $\bmat{ccc} \! \! \!
-P_{65}'^4 & \, P_{16}'^2 & \, P_{51}'^3 \\
\, P_{62}'^1 & \! \! \! -P_{46}'^5 & \, P_{24}'^3 \\
\, P_{35}'^1 & \, P_{43}'^2 & \! \! \! -P_{54}'^6 \emat $ & $\bmat{ccc} g_{11}'
& g_{12}' & g_{13}' \\
g_{21}' & g_{22}' & g_{23}' \\
g_{31}' & g_{32}' & g_{33}' \emat$&$\begin{matrix}g'_{ii}=p'_6\\
g'_{12}=g'_{23}=g'_{31}=p'_3\\g'_{13}=g'_{21}=g'_{32}=-p'_2\end{matrix}$
 \\[0.45cm]

 $\bmat{ccc} \! \! \! P_{32}'^4 &
P_{13}'^5 & P_{21}'^6  \! \! \! \emat$ & $ \! \! -\bmat{ccc} \! \!
\!  \bar f_1' & \bar f_2' & \bar f_3'  \! \!
\! \emat$ &$-p'_1$ \\[0.35cm]

 $\bmat{ccc} \!
\! \!
-P_{23}'^1 & \, P_{34}'^5 & \, P_{42}'^6 \\
\, P_{53}'^4 & \! \! \! -P_{31}'^2 & \, P_{15}'^6 \\
\, P_{26}'^4 & \, P_{61}'^5 & \! \! \! -P_{12}'^3 \emat$ & $\bmat{ccc}
\bar g_{11}' & \bar g_{12}' & \bar g_{13}' \\
\bar g_{21}' & \bar g_{22}' & \bar g_{23}' \\
\bar g_{31}' & \bar g_{32}' & \bar g_{33}' \emat$&$\begin{matrix}
\bar g'_{ii}=-p'_0\\
\bar g'_{12}=\bar g'_{23}=\bar g'_{31}=p'_4\\
\bar g'_{13}=\bar g'_{21}=\bar g'_{32}=-p'_5\end{matrix}$
\end{tabular}


\end{center}

\label{Dicccionario1}

\end{table}

\end{document}